\documentclass[letterpaper,11pt]{article}
\usepackage{jinstpub} % for details on the use of the package, please see the JINST-author-manual
\usepackage{lineno}
\usepackage{subcaption}
\usepackage{multirow}
\usepackage{cleveref}

\title{\boldmath Ultra-Fast Generation of Air Shower Images for Imaging Air Cherenkov Telescopes using Generative Adversarial Networks}
\author{Christian Elflein$^*$, Stefan Funk$^*$, Jonas Glombitza$^*$}
\affiliation[*]{Friedrich-Alexander-Universit\"at Erlangen-N\"urnberg, Erlangen Centre for Astroparticle Physics, Nikolaus-Fiebiger-Str. 2, 91058 Erlangen, Germany}

\emailAdd{christian.elflein@fau.de}
\emailAdd{jonas.glombitza@fau.de}

\abstract{For the analysis of data taken by Imaging Air Cherenkov Telescopes (IACTs), a large number of air shower simulations are needed to derive the instrument response. The simulations are very complex, involving computational and memory-intensive calculations, and are usually performed repeatedly for different observation intervals to take into account the varying optical sensitivity of the instrument.
The use of generative models based on deep neural networks offers the prospect for memory-efficient storing of huge simulation libraries and cost-effective generation of a large number of simulations in an extremely short time.
In this work, we use Wasserstein Generative Adversarial Networks to generate photon showers for an IACT equipped with the FlashCam design, which has more than 1,500 pixels.
Using simulations of the H.E.S.S. experiment, we demonstrate the successful generation of high-quality IACT images.
The analysis includes a comprehensive study of the generated image quality based on low-level observables and the well-known Hillas parameters that describe the shower shape.
We demonstrate for the first time that the generated images have high fidelity with respect to low-level observables, the Hillas parameters, their physical properties, as well as their correlations.
The found increase in generation speed in the order of $10^5$ yields promising prospects for fast and memory-efficient simulations of air showers for IACTs.}


\arxivnumber{2311.01385} % Only if you have one

\dedicated{\flushleft\textnormal{\textsc{%
Published in JINST as \href{https://iopscience.iop.org/article/10.1088/1748-0221/19/04/P04010}{DOI:10.1088/1748-0221/19/04/P04010}
\\
}}}

\makeatletter
\gdef\@fpheader{}
\makeatother

\begin{document}
\maketitle
\flushbottom

\section{Introduction}
\label{sec:intro}
In the last two decades, gamma-ray observatories, such as arrays of Imaging Air Cherenkov Telescopes (IACTs), have opened a new window to the non-thermal universe by providing accurate observations of the very-high-energy (VHE) gamma-ray sky.
The High Energy Stereoscopic System (H.E.S.S.)~\cite{hess_crab}, located in Namibia, is a system of five IACTs investigating cosmic gamma rays in the GeV and TeV regimes.
Using IACTs, Cherenkov radiation emitted from secondary particles in extensive air showers (EAS) initiated by cosmic particles penetrating the Earth's atmosphere can be detected. The emitted Cherenkov radiation is imaged using the cameras of the IACTs, and subsequently, information about the primary particle is gained by analyzing the obtained camera images.
Traditionally, the so-called Hillas parameters~\cite{hillas1985cerenkov} are utilized to describe the image by modeling the Cherenkov light in the camera as moments of the light-intensity distribution.
While the major axis is used to reconstruct the shower axis, i.e., the arrival direction of the incident particle, the \emph{size}, i.e., the integrated intensity of the image, is used to estimate the shower's energy.
The shape of the distribution, i.e., its width and length, as well as higher moments of the Cherenkov light distribution, are commonly used to determine the species of the primary particle~\cite{OHM2009383}.

To perform detailed gamma-ray observations, the instrument and its performance have to be well understood.
Such studies require a large number of Monte Carlo (MC) simulations, in particular, the most costly to produce being of the hadronic background.
In the astroparticle physics community, the standard for the simulation of air showers and the Cherenkov emission is the software program CORSIKA (COsmic Ray SImulations for KAscade)~\cite{heck_corsika_1998}.
The instrument response of IACTs is commonly simulated using the program package \texttt{sim\_telarray}~\cite{simtel}.
Like in measurements under real operation conditions, camera images containing information about the primary particle are obtained after running the full simulation chain.
Running these simulations is computationally expensive, as in the order of $10^6$ events, or more are needed due to the imbalance in the ratio of photon showers (signal) and hadron showers (background) of 1 to a few thousand, and the simulation time of a single event is in the order of $1-10$~s (depending on the energy).
Due to aging effects of the instrument for different run periods, new simulations have to be produced, in which the changes are usually small.
Furthermore, storing these comprehensive libraries is rather expensive and typically in the order of hundreds of Gigabytes for one observation phase.

The recent advances in unsupervised machine learning using neural networks provide new techniques to approximate high-dimensional probability densities using generative models and apply them to various physics challenges~\cite{dlfpr}.
In particle physics, a large community is investigating the use of Generative Adversarial Networks (GAN)~\cite{Paganini:2017dwg, Erdmann:2018jxd, Chekalina:2018hxi, Carminati:2018khv, Musella:2018rdi, Deja:2019vcv, SHiP:2019gcl, Butter:2019cae, hashemi2023ultrahighresolution}, as well as normalizing flows~\cite{Krause_2023,krause2023caloflow} or diffusion models~\cite{buhmann2023caloclouds} to accelerate the simulation of calorimeter data and deal with the computational requirements in high-energy physics.
In particular, GANs and normalizing flows offer the possibility to accelerate the generation of new simulations by orders of $10^4$ to $10^5$ without large losses in the physical quality. Diffusion models seem to provide very high fidelity but currently seem to lack speed and computational efficiency.

Also, in the field of astroparticle physics, the applications are broad. For example, only a fraction of simulations would have to be simulated; with the help of generative models, the statistics could then be increased or the simulations refined using an adversarial set-up~\cite{Erdmann:2018kuh}.
Likewise, the simulation under slightly different conditions could be performed with only a small additional effort since less data is needed to retrain the generative model by utilizing transfer learning~\cite{wang2018transferring}.
In addition, the models would allow the generation of events conditioned on parameters that are not directly accessible with the MC codes and would require costly calculations, e.g., simulating proton showers with a shower maximum distribution uniform in $X_\mathrm{max}$ to check systematics of the instruments.

Since the simulation includes both the physics simulations and the detector response, the optimal treatment for such an approach would be to use two models.
As an initial step, we will concentrate on a simplified approach, which is commonly pursued in event generation using generative models that combine both parts in a single model.

The computational requirements for the next-generation flagship in gamma astronomy, the Cherenkov Telescope Array (CTA)~\cite{CTA}, led to the first promising studies of fast, memory-efficient, and sustainable generation of IACT images using generative models~\cite{taiga_1, taiga_2, veritas_wgan, wgan_veritas} based on conditional GANs~\cite{odena2017conditional} or Wasserstein GANs~\cite{gulrajani2017improved}.
Despite showing encouraging results in generating the high-level image parameters directly~\cite{tactic}, none of them was, so far, able to provide high-resolution IACT images with high quality, including a comprehensive and convincing analysis of the fidelity of the generated images, e.g., by studying Hillas parameters and their correlations.

In this work, we adopt a Wasserstein GAN to perform the generation of IACT air shower images.
As a proof of concept, we use simulations of the H.E.S.S. CT5 telescope with its new camera having more than 1,500 pixels that utilize the FlashCam design~\cite{puehlhofer2021science} foreseen for the medium-sized telescopes of CTA.
The integration of stereoscopy into deep-learning-based algorithms is an ongoing challenge in the community, even for reconstruction tasks~\cite{Shilon_2019, Brill_2019, ct_learn, Spencer_2021, Glombitza_2023, Glombitza:20234C} and many works are, consequently, currently focusing on single-telescope reconstructions~\cite{nieto_hex, jacquemont:hal-03043188, Jacquemont_2021}.
Similar to previous studies including generative models in the context of IACTs, we, thus, consider the implementation of image stereoscopy as a study to be, for the moment, disentangled from the challenge of fast simulations.
We show for the first time the successful generation of IACT images with high quality and demonstrate that the generated images feature high fidelity in terms of the encoded physics parameters as well as the Hillas parameters and their correlations.

\section{Experimental setup}
\label{sec:setup}

In this work, we utilize H.E.S.S. simulations; however, our developed technique can be applied to any Cherenkov telescope with a regular pixel placement.
Four of the five H.E.S.S. telescopes (CT1-4) are arranged in a square with a side length of 120~m, and the fifth larger telescope (CT5) was added in the center of the square array in 2012.
H.E.S.S. features different observation modes~\cite{parsons2015hess}, each characterized by the number of operating telescopes. We limit ourselves to \textit{mono} observations, in which events are only read out if the trigger criteria in CT5 are met.
The remaining modes are called \textit{stereo} and \textit{hybrid} and feature CT1-4 and CT1-5 in the observation, respectively.
The generation of events, including multiple telescopes, i.e., stereoscopic events, is left open for future work.
Nonetheless, we address another challenge in our work, as the CT5 camera's resolution with 1764 pixels is significantly higher than for the other telescopes CT1-4.
Furthermore, note that by choosing CT5-only events, we apply our algorithm to the FlashCam design that is also foreseen in CTA.

\subsection{Reference dataset}
The data used for the image generation in this work were simulated with CORSIKA and \texttt{sim\_telarray}.
During the simulation of the air shower, EGS4~\cite{nelson1990egs4} and QGSJET-II-04~\cite{Ostapchenko_2014} were used for simulating the electromagnetic and hadronic particle interactions, respectively.
We used a zenith angle of $20^{\circ}$, an opening angle of $5^{\circ}$, and an energy range of $10^{-1.5}$~GeV to $10^2$~TeV, with a spectral index of $E^{-2}$ for the simulation of the air showers.
After obtaining the raw data, a calibration process was carried out, in which the measured analog-to-digital converter (ADC) counts were converted into the physical unit of photoelectrons (p.e.).
After the calibration, image clearing was applied to remove the night sky background (NSB) so that mostly the signal pixels are kept in the image.
The so-called tail cuts cleaning process retains pixels only if their signal values exceed a specified upper threshold, while at least one neighboring pixel value exceeds the lower threshold, and conversely.
In this work, we performed extended 4/7 cleaning, i.e., tail cuts cleaning is applied with 7 p.e. and 4 p.e. as thresholds, but four rows around the cleaned signal pixels are kept to `extend' the image.
The extended cleaning using lower thresholds\footnote{Commonly for CT5 9/16 cleaning is applied.} is usually used within H.E.S.S. to include low signals in the camera image, i.e., the detailed shower substructure.
The generation of NSB-only pixels is not considered in this work and is left for the future.
Lastly, in our study, we only consider images with an integrated signal larger than a 250 p.e. signal, which is used as a standard cut in the mono analysis \cite{pühlhofer2021science}.
The final data set for this study consists of a total of ${\sim}515,000$ images.
We use ${\sim}361,000$ for training, ${\sim}77,000$ for validation, and ${\sim}77,000$ for the final test of the image quality.

\subsection{Pre-processing of the data}

Next, the data is carefully prepared to guarantee a stable training of the generative model.
%First, the data is split into training, validation, and test data sets with a ratio of 0.7:0.15:0.15.
For this work, we considered two physical properties of the air shower images to be conditioned during the image generation.
These properties are the energy \(E\) of the primary particle and air shower impact point \(I = (I_x, I_y)\) defined by the intersection of the ground and the shower axis.

\paragraph{IACT images}
Pixels with negative values are set to zero, and the logarithmic transformation $\log_{10}(1+s_i)$ is applied to all pixel values $s_i$.
We further normalize all pixel values by dividing them by the standard deviation estimated over all pixel values of the training data set.
Since our algorithm relies on a Cartesian structure, a transformation from the hexagonal pixel placement into a Cartesian representation is needed~\cite{bryan_kim_2023_7982088}.
In this work, axial indexing~\cite{ERDMANN201846} is applied to the initial camera images, resulting in Cartesian images of dimension 56 \(\times\) 56.

\paragraph{Energy and impact point}

To improve the stability of the neural network training, the energy is transformed to $\log(E/\text{TeV})$ and normalized to [0, 1].
In the case of the impact point, both coordinates are standardized individually using z-score normalization, i.e., to a distribution with zero mean and unit variance.

\section{Ultra-fast image generation using Generative Adversarial Networks}

In recent years, the field of generative models has been continuously expanding, and several types of generative models \cite{ruthotto2021introduction} showed impressive results in the generation of natural images~\cite{dhariwal2021diffusion, brock2019large, karras2019stylebased, karras2020analyzing}.
Particularly, generative adversarial networks (GANs)~\cite{Goodfellow:2014upx} have enjoyed great popularity.

\subsection{Generative Adversarial Networks (GANs)}

GANs are a particular generative model consisting of two networks trained in an adversarial manner.
The two networks are usually called generator \(G\) and discriminator \(D\).
The generator tries to approximate the real data distribution \(p_r\) with samples \(x_r\) using a distribution \(p_{\theta}\) and its samples $x_\theta$.
Whereas $\theta$ denotes the adaptive parameters of the network.
It takes noise \(z\) from a simple distribution \(p_z\), which is usually a multivariate Gaussian, as input and learns to map this input to new samples \(x_{\theta} = G_{\theta}(z)\), similar to real samples.
To improve the generator performance, the second network, the discriminator, takes samples from both the real data and model distribution and outputs the probability of the samples being fake or real.
This yields the following cross-entropy-like loss, commonly used in the supervised training of classifiers, for GAN training
\begin{equation}
\label{eq:gan_loss}
\begin{aligned}
\mathcal{L}_{\text{GAN}} = \min_G \max_D \; \mathbb{E}_{x_r \sim p_r}\left[\log(D(x_r))\right] + \mathbb{E}_{z\sim p_z}\left[\log(1-D(G_{\theta}(z)))\right]
\end{aligned}
\end{equation}
in which $\mathbb{E}$ denotes the expectation value estimated using the respective batch of data samples.
During the discriminator training, the network aims to learn the discrimination between real $x_r$ and generated samples $x_\theta$ by maximizing \(\mathcal{L}_{\text{GAN}}\) similar to training a classifier in a supervised fashion.
A perfect discriminator would output \(D(x_r) = 1\) when evaluating real samples and \(D(G_{\theta}(z)) = 0\) when evaluating fake samples produced by the generator, which would result in the maximum of $\mathcal{L}_\mathrm{GAN} = 0$.
In contrast, during the training of the generator, the loss is minimized to fool the discriminator and produce realistic-looking samples, i.e., update the generator in such a way that $D(G_{\theta}(z))\rightarrow 1$.
Due to the adversarial training of GANs, both networks continuously improve throughout the training.
To not disturb the training of the networks and ensure reasonable updates, during the generator update, the adaptive parameters of the discriminator are fixed and during the discriminator training, the parameters of the generator are fixed. 

\subsection{Wasserstein Generative Adversarial Networks}

To overcome the challenges of GANs~\cite{arjovsky2017principled}, which are cumbersome to train and suffer from mode collapse, i.e., generate samples from a restricted phase space only, in this work, the improved version~\cite{gulrajani2017improved} of the Wasserstein GAN \cite{arjovsky2017wasserstein} is used.
The significant difference is that the discriminator no longer outputs a probability.
Instead, the network takes samples \(x\) from the real data and the model distribution and estimates the Wasserstein distance.
The Wasserstein distance, which is also called Earth-Mover's distance, is a measure of the similarity of the distributions and describes the minimal amount of work needed to transform one distribution into another.
The discriminator is, therefore, often called critic \(C\) since it does not classify between real and generated in the WGAN but instead gives feedback about the similarity of the distributions.

Using the Kantorovich-Rubinstein duality
\begin{equation}
\label{eq:wasserstein_distance}
\begin{aligned}
W = \underset{\left\Vert f \right\Vert _L\geq 1}{\text{sup}} \;\mathbb{E}_{x_r \sim p_r}\left[f(x_r)\right] - \mathbb{E}_{x_{\theta} \sim p_{\theta}}\left[f(x_{\theta})\right]
\end{aligned}
\end{equation}
the Wasserstein distance can be estimated in the WGAN training by approximating $f(x)$ as the critic network C.
Therefore, the Lipschitz condition is constrained by training the critic to be a 1-Lipschitz function using the gradient penalty 
\begin{equation}
\label{eq:gradient_penalty}
\begin{aligned}
\mathcal{L}_{\mathrm{GP}} = \lambda \cdot \mathbb{E}_{x_u\sim p_u}\left[\left(\lVert \nabla_{x_u} C(x_u)\rVert _2 - 1\right)^2 \right],
\end{aligned}
\end{equation}
which penalizes the gradient on a mixture sample $x_u$ defined via:
\begin{equation}
\label{eq:uniform_sample}
\begin{aligned}
x_u = \epsilon \cdot x_r + (1-\epsilon) \cdot x_{\theta}
\end{aligned}
\end{equation}
to ensure a critic gradient with norm 1, where the parameter \(\epsilon\) is a random number sampled from a uniform distribution \(U[0, 1]\).
This ensures well-behaved critic gradients even if the critic is evaluated outside the support of $p_r$ and $p_\theta$, and thus, informative updates for the generator during the WGAN training.
Combining equation~\ref{eq:wasserstein_distance} and \ref{eq:gradient_penalty} results in the loss for the WGAN with gradient penalty used in this work:
\begin{equation}
\label{eq:wgan_loss}
\begin{aligned}
\mathcal{L}_{\text{WGAN-GP}} = \mathbb{E}_{x_r\sim p_r}[C(x_r)] - \mathbb{E}_{z\sim p_z}[C(G_{\theta}(z))] + \lambda\,\mathbb{E}_{x_u\sim p_u}[(\left\lVert \nabla C(x_u)\right\rVert _2 - 1)^2].
\end{aligned}
\end{equation}
The main benefits of this change are the possible training of the critic to convergence, i.e., performing $n_\text{critic}$ updates before updating the generator, the reduction of mode collapsing, and an objective that correlates with the quality of the generated samples.
In WGAN training, the critic is usually trained several times before each generator update to gain a proper estimation of the Wasserstein distance and, thus, precise feedback.
Due to the improved properties of WGANs and their successful application in physics~\cite{Erdmann:2018kuh, wgan_veritas, Erdmann:2018jxd, Chekalina:2018hxi, Buhmann_2021, Di_Sipio_2019, atlas_8588710}, the sophisticated algorithm will be used in the following to generate air shower camera images.

\subsubsection{Conditioning of labels}

To design a generator able to control the properties of generated samples, i.e., the generation of events with a specific energy \(E\) of the primary particle and the air shower impact point \(I = (I_x, I_y)\) that is the distance of the shower to the telescope, label conditioning is used~\cite{odena2017conditional}.
For that, two constrainer networks are added to the WGAN framework, one for each class label.
These networks take generated images \(x_\theta\) as input, reconstruct the corresponding physical property, and provide feedback to the generator to ensure the labels given to the generator are correctly represented in the images.
To further improve the reconstruction, the class label not reconstructed in the respective constrainer networks is used as an additional input; e.g., the primary particle energy is used to reconstruct the impact point better.

The mean-squared-error objective function is used for both constrainer networks and reads 
\begin{equation}
\label{eq:const_loss}
\begin{aligned}
%\mathcal{L}_{\text{con, i}} = [y_i - c_i(x, y_j)]^2
\mathcal{L}_{\text{con, E}} = [E - \hat{E}(x, I)]^2\\
\mathcal{L}_{\text{con, I}} = [I - \hat{I}(x, E)]^2 
\end{aligned}
\end{equation}
for the two networks $\hat{E}$ and $\hat{I}$.
Both constrainer networks are trained by minimizing Eq.~\ref{eq:const_loss} in parallel to the WGAN training procedure after each generator iteration using the real samples $x_r$.

To enforce the correct implementation of the physical properties in the generation process, the class labels are given to the generator for generating images \(x_\theta = G_{\theta}(z, E, I)\) with specific energies and impact.
And the auxiliary loss:
\begin{equation}
\label{eq:aux_loss}
\begin{aligned}
%\mathcal{L}_{\text{aux}} = \sum_i^n \alpha_i mathcal{L}_{\text{con, i, gen}}
%\mathcal{L}_{\text{aux}} = \alpha_i \mathbb{E}_{x_\theta \sim p_\theta, I\sim p_I, E\sim p_E} [\mathcal{L}_{\text{con}(x_\theta, E, I)}]
\mathcal{L}_{\text{aux}} = \alpha_I \cdot \mathcal{L}_{\text{con, I}}(x_\theta, E, I) + \alpha_E \cdot \mathcal{L}_{\text{con, E}}(x_\theta, E,I) 
\end{aligned}
\end{equation}
is added to the default generator loss and minimized during the generator update.
The hyperparameters $\alpha$ are used to control the size of the constraint by scaling the losses.
Similar to the critic, the parameters of the constrainer networks are fixed in the generator update step.

Lastly, the class labels are also given to the critic to improve the estimation of the Wasserstein distance conditioned to the respective label during the critic training and, in turn, improve the feedback for the generator.

\subsection{Design of the adversarial framework and training strategy}
\label{sec:training_strategy}

The framework for generating IACT images consists of the four different neural networks described above: the generator, the critic, and the two constrainer networks.
The generator is based on transposed convolutions and receives a 100-dimensional noise vector, the primary particle energy $E$, and the air shower impact point $I$ as input. The output is an IACT image $x_\theta$ with dimension $56 \times 56$.
As the last layer of the network, we use the rectified linear unit (ReLU)~\cite{pmlr-v15-glorot11a} activation function $\sigma_{\text{ReLU}} = \text{max}(0, x)$ to guarantee generated signals larger than 0.
The critic is a convolutional neural network (CNN)~\cite{lecun1998gradient} that takes simulated and generated images and their corresponding class labels as input and estimates the Wasserstein distance.
Also, the two constrainer networks are CNNs.
To ensure precise feedback, the CNNs feature leaky ReLU activations and residual blocks~\cite{he2015deep}.
The exact designs of all networks is summarized in Table~\ref{tab:gen_arch}, Table~\ref{tab:critic_arch}, Table~\ref{tab:energy_arch}, and Table~\ref{tab:impact_arch}.

The images and labels of the training data set are used as the real events for the WGAN training to ensure an accurate correlation between energy and impact point.
The validation data set is used to observe the potential overtraining of the constrainer networks, the test data set is used for the final analysis of the WGAN.
The generated images are obtained by sampling noise from a $100$-dimensional multivariate Gaussian and labels from the energy and impact distribution after applying Gaussian smearing and feeding them into the generator.
The energy labels are smeared by 10\,\%, and the impact labels by 10\,m.
These smeared training labels for generating images are chosen due to the correlation of the image with energy and impact point, which would not be kept if the physical labels were sampled separately.
Replacing this approach, using, for example, normalizing flows, is outside the scope of this work but planned for the future.

The WGAN was trained for 1000 epochs on an NVIDIA A100-SXM4-40GB, which took about 62 hours.
In one epoch, the WGAN iterates over the whole training data.
We aim to split the training set into as large batches as possible to obtain the most precise feedback from the discriminator~\cite{brock2019large}. Thus, we used a batch size of 512, close to the VRAM limit of our GPU. 
The critic was trained using all batches, and in every \(n_{\text{critic}} = 10\) batches~\cite{gulrajani2017improved}, the generator, and the constrainer networks were trained. 
The weight of the gradient penalty was set to \(\lambda = 10\)~\cite{gulrajani2017improved}.
For the generator and the critic, the Adam optimizer~\cite{kingma_adam_2017} with an initial learning rate of \(lr = 10^{-4}\), \(\beta_1 = 0.5\), and \(\beta_2 = 0.9\) was used, which was adapted from Ref.~\cite{gulrajani2017improved}.
Both of the constrainer networks used Adam with \(lr = 10^{-3}\), \(\beta_1 = 0.5\), and \(\beta_2 = 0.5\).
The auxiliary parameters for scaling the constrainer losses were set to \(\alpha_E = 500\) and \(\alpha_I = 50\).
The specific parameter ratio is selected to achieve an equal contribution of both constrainer losses to the auxiliary loss.
The order of magnitude of these parameters is chosen so that the generator is conditioned properly.
The exact values were determined through manual hyperparameter tuning.
The learning rate of all networks was decreased exponentially by about an order of magnitude during the training.

\subsection{Computational speed-up}

Among the various advantages of deep-learning-based image generation, the most outstanding one is the computational speed-up.
The computation times for standard simulation and the image generation with our framework are presented in \Cref{tab:gen_speed}.
While the simulation of 100,000 IACT images takes almost three days with CORSIKA and \texttt{sim\_telarray}, the WGAN generates the same amount of images within roughly a minute on an approximately competitive computing architecture, translating to a speed-up of roughly 3000.
Parallelizing the generation of images with the WGAN on a graphics processing unit (GPU) results in a computational speed-up of more than five orders of magnitude.

\begin{table}[t!]
\centering
\begin{centering}
\begin{tabular}{ c | c | c | c }
Method & Hardware & Time & \textbf{Speed-up} \\
\hline
\hline
CORSIKA \& \texttt{sim\_telarray} & Intel Xeon Gold 6230 & 70h & -- \\
WGAN framework & AMD EPYC 7713 Milan & 86.06\,s & \textbf{x\,2930} \\
WGAN framework & NVIDIA A100-SXM4-40GB & 2.34\,s & \textbf{x\,108000} \\
\hline
\end{tabular}
\caption{Computational time required for generating 100,000 IACT images using different methods.}
\label{tab:gen_speed}
\end{centering}
\end{table}

\section{Visual inspection and analysis of low-level characteristics}

\begin{figure}[t!]
\centering
\includegraphics[width=1.\textwidth]{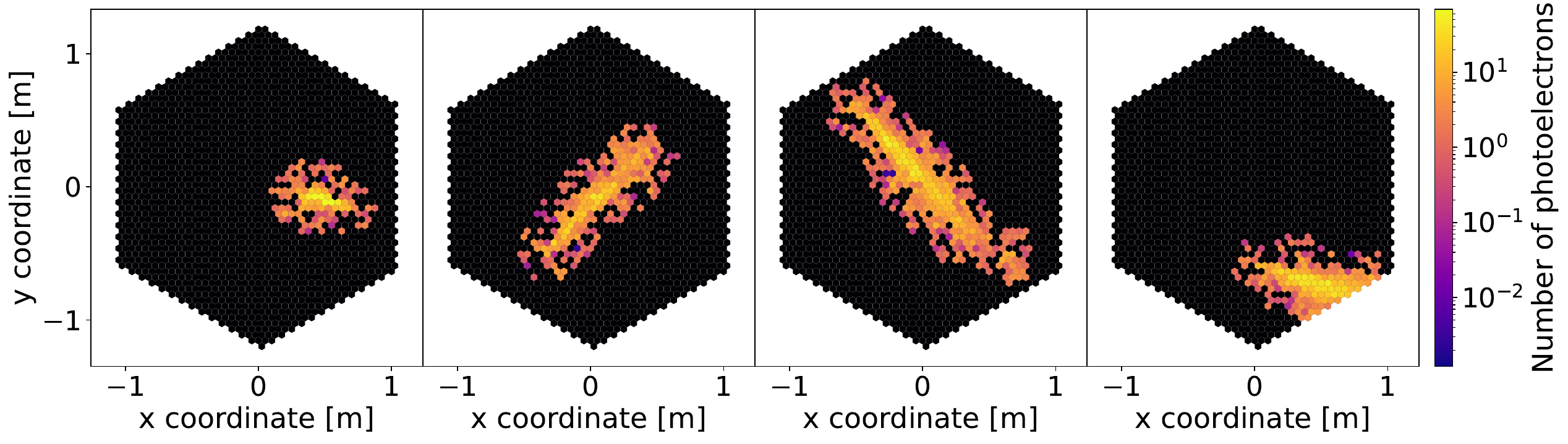}
\qquad
\includegraphics[width=1\textwidth]{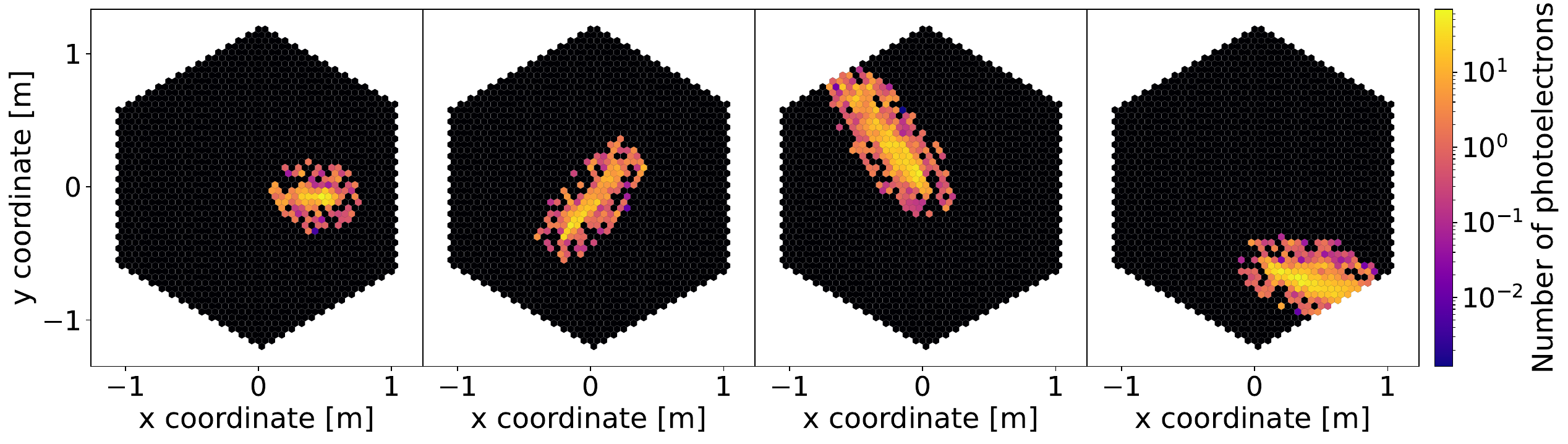}
\caption{Four IACT images from the simulated test data set (top) compared to four images from the generated data set (bottom). The simulated images were handpicked to find samples that feature similar characteristics and are compared to the most similar samples in the generated data set.
Pixels that do not contain signals are shown in black.
\label{fig:real_fake_images}}
\end{figure}

In the following section, the quality of the generated images is examined using various techniques.
For these comparisons, we make use of the test data set --- not used during training --- and generate samples using our WGAN conditioned on the label distributions of the energy and impact point of the reference data set.
In Figure~\ref{fig:real_fake_images}, we present four representative types of showers in the data set --- specifically chosen to highlight the different shapes typically represented in IACT images --- and compare them to similar images of the generated samples, whereas as generated images, the sample with the smallest pixel-wise MSE w.r.t. the test image was selected out of the generated data set.
In the first column, the signals in both images resemble a circular shape, corresponding to a shower developing in parallel to the telescope's line of sight with a close-by impact point.
Images with the typical elliptical-shaped signals can be seen in the second and third columns.
A representative image for so-called truncated images, in which the Cherenkov light falls to the edge of the telescope's field of view, is shown in the last column.
These samples represent a large variety of images, which could be successfully generated using the WGAN. Additionally, for an unbiased comparison, randomly chosen samples from the test data set and the generated samples can be found in the appendix in Figure~\ref{fig:add_images}.

After inspecting the visual image quality, a comprehensive investigation of low-level parameters is performed to determine the quality quantitatively.
As a first step, we examine how well low-level parameters are resembled in the generated images before studying high-level parameters, their correlations, and dependency on the label conditioning in Sec.~\ref{sec:high_level}.
The three studied low-level parameters that need to be precisely modeled are the pixel occupancy, the number of triggered pixels, and pixel intensities.

\begin{figure}[t]
\centering
\includegraphics[width=0.95\textwidth]{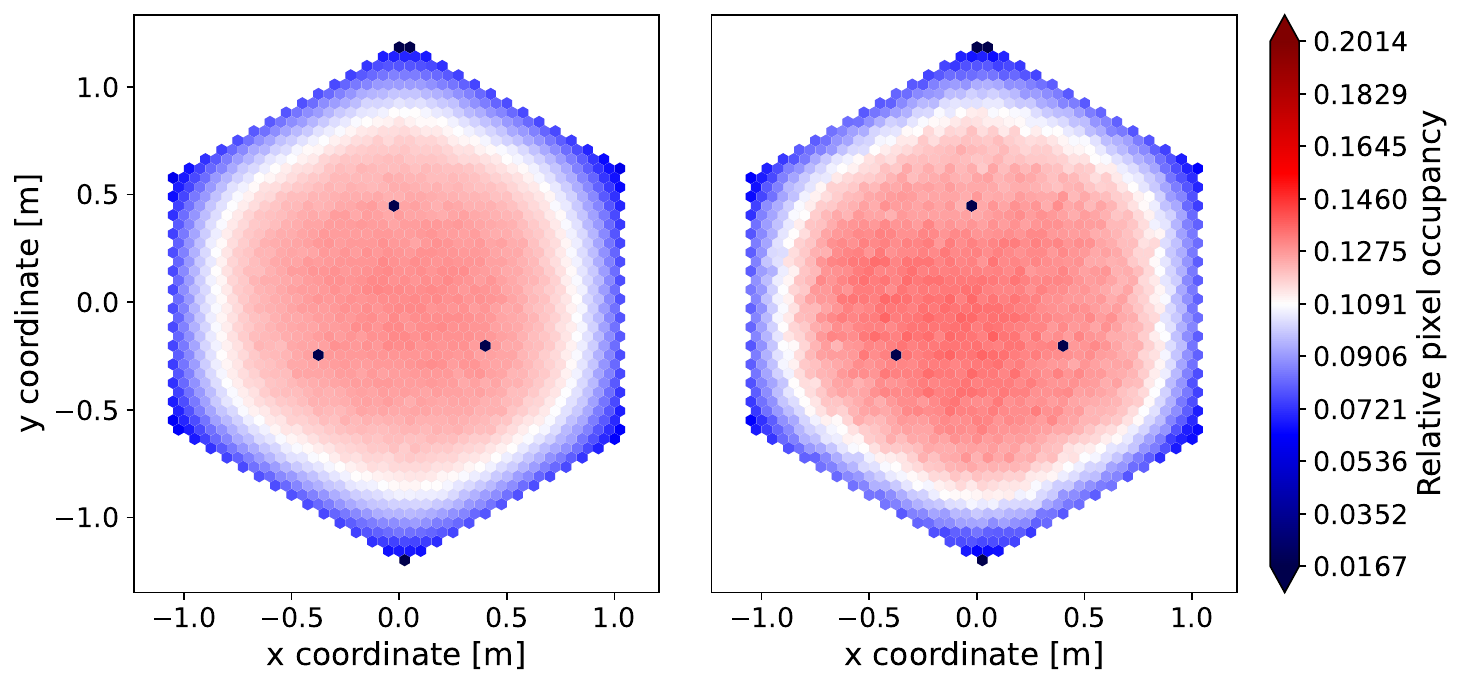}
\caption{Camera images displaying the relative pixel occupancy for the simulated data set (left) and the generated data set (right). The colorbar shows the \(5\sigma\) range of the mean occupancy value of the simulated mean image.\label{fig:occupancy_image}}
\end{figure}

The pixel occupancy, i.e., how often a pixel holds signals averaged over the whole dataset, is shown in Figure~\ref{fig:occupancy_image}, where the color bar ranges from -5 to 5 $\sigma$ deviation (statistical).
Overall, the pixel occupancies of the simulated and generated data sets are very similar and show the same behavior of a smaller occupancy at the edges of the camera and a larger occupancy in the central area of the camera.
Combined with the finding of the image quality, these findings show that our generative algorithm does not suffer from mode collapsing, as present in previous works \cite{taiga_1, taiga_2}.
The three vertically outermost pixels in the simulated image, as well as the mount pixels, i.e., the three triangularly-arranged pixels in the center of the camera, are not simulated in its current design. In this approach, we, therefore, are not considering the pixels and they are masked to zero in the last layer of the generator network.
Comparing both plots, the generated image is more dispersed, which is especially shown in the middle of the camera. However, the fluctuations are, in comparison to the overall gradient from the outside to the inside of the camera, small.

\begin{figure}[t]
    \begin{subfigure}[b]{0.5\textwidth}
        \centering
        \includegraphics[width=.99\textwidth]{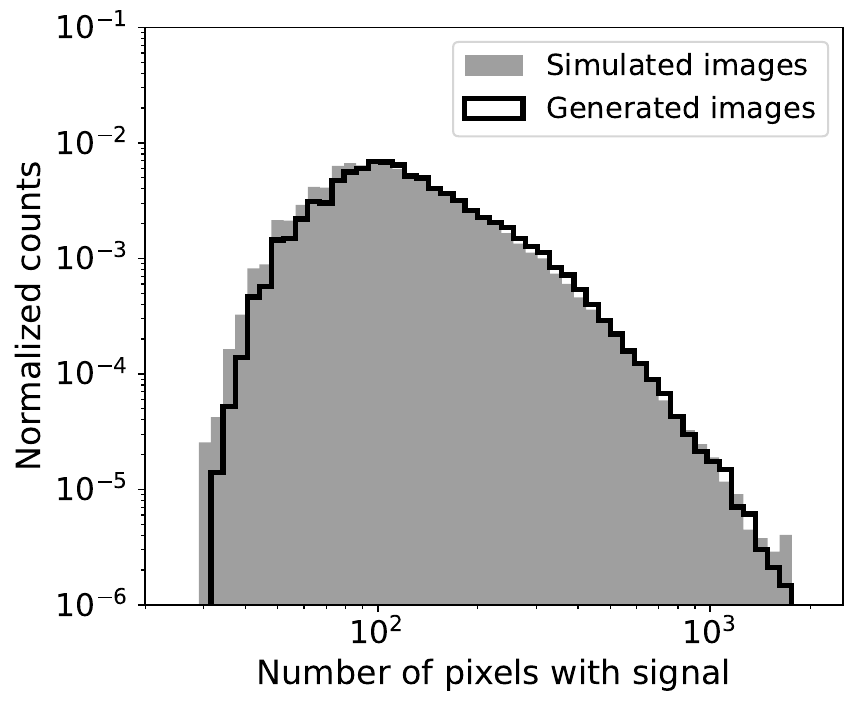}
        \caption{}
        \label{fig:nr_pixels}
    \end{subfigure}
    \begin{subfigure}[b]{0.5\textwidth}
        \centering
        \includegraphics[width=.99\textwidth]{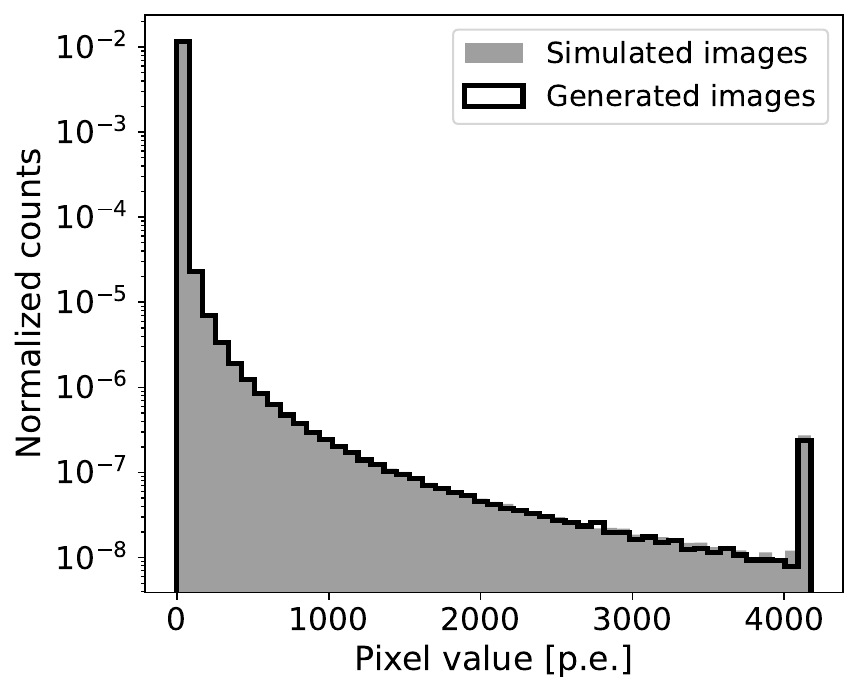}
        \caption{}
        \label{fig:pixel_values}
    \end{subfigure}
    \caption{Distributions of the (a) number of pixels per image containing signals and (b) pixel values for generated (black) and simulated (grey) events.\label{fig:nr_pixels_and_values}}
\end{figure}

The distribution of triggered pixels per event is shown in Figure~\ref{fig:nr_pixels}.
For this and all the following plots, we use the same style, in which the simulation is shown as a colored grey histogram, and the generated histogram is shown as a solid black line.
The distributions of the number of signal pixels per image follow the same trend for both simulated and generated images and show a good agreement.
The most occurring number of signal pixels is around 100, with a decreasing trend to lower and higher values.
The cut-off of the distributions at 1,758 is due to the number of pixels of the CT5 camera.
The most noticeable but still moderate difference between the distributions can be seen for images with fewer signal pixels, where the generated data set features fewer images with few signal pixels.
Nevertheless, the overall match is good.

In Fig.~\ref{fig:pixel_values}, the distributions of pixel values are compared for both data sets.
Pixels without signals are included in the smallest bin.
Both distributions match well over the whole range up to the limit at $\sim$ 4200~p.e. caused by saturation.
Overall, we find a good agreement between the simulated test and the generated data set for the studied low-level parameters over the entire range, which covers several orders of magnitude.

\section{\label{sec:high_level}Analysis of high-level parameters}
After comparing low-level parameters, the comparison of high-level parameters is discussed below.
Prior to investigating image parameters in section~\ref{sec:hillas_parameters}, the label-conditioning of the generation process is evaluated by comparing the injected air-shower property (energy, impact) with the reconstructed event property using the corresponding constrainer networks.
Lastly, in section~\ref{sec:parameter_correlations}, the correlation between the high-level parameters is investigated to check how accurately the WGAN can model the complex relations.

\subsection{Label conditioning: shower energy and impact point}
\label{sec:energy_and_impact}

Verifying the correct implementation of physical properties in the generated images is essential.
For the IACT image generation in this work, two physical labels, namely the primary particle energy \(E\) and the air shower impact point \(I = (I_x, I_y)\), are used.
For generating new images, randomly sampled noise and these physical labels are input into the generator.
During training, the energy and impact constrainer networks --- trained to reconstruct the respective label using simulated data only ---  are utilized to enforce the label representation in the final image.
Using the constrainer networks, the physical labels of the generated images can be reconstructed.
Comparing the reconstructed labels with the labels initially given into the generator provides information about the implementation of the physical labels into the generated samples.

\begin{figure}[t!]
    \centering
    \begin{subfigure}[b]{0.45\textwidth}
        \centering
        \includegraphics[width=.99\textwidth]{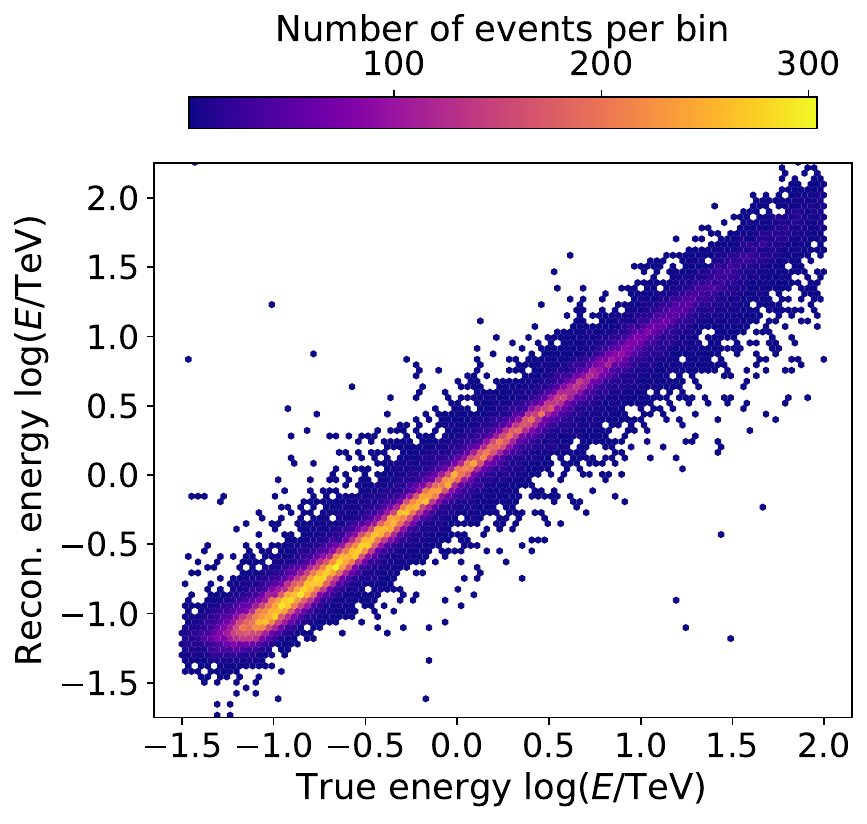}
        \caption{}
        \label{fig:rec_constrainer_mc}
    \end{subfigure}
    \begin{subfigure}[b]{0.45\textwidth}
        \centering
        \includegraphics[width=.99\textwidth]{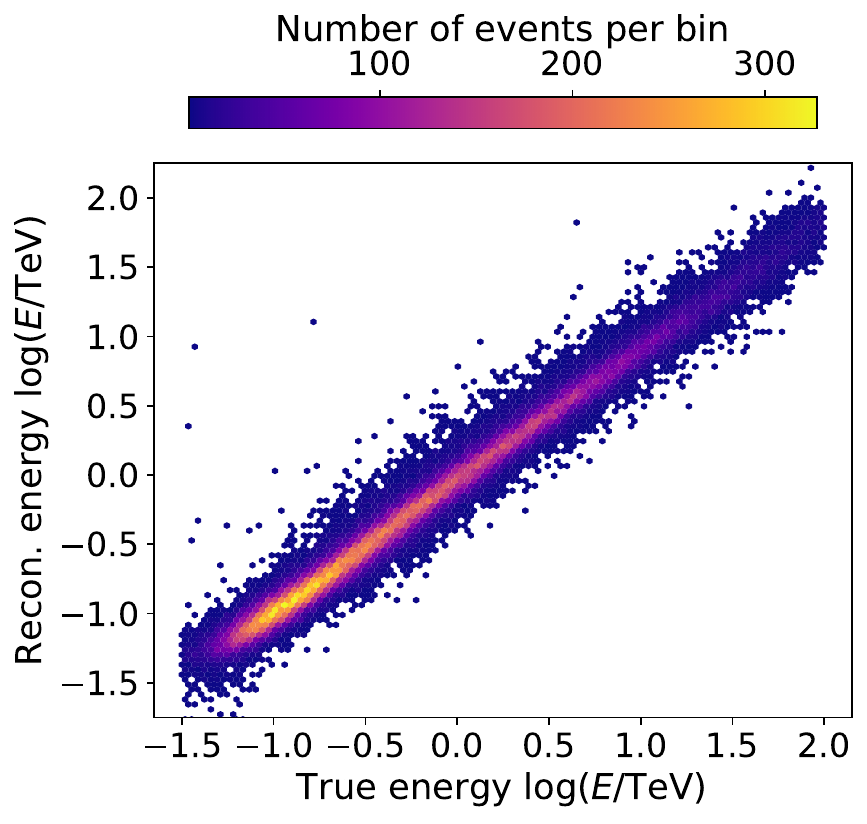}
        \caption{}
        \label{fig:rec_constrainer_gen}
    \end{subfigure}
\caption{Reconstruction of the primary particle energy \(E\) using the energy constrainer network for (a) simulated events and (b) generated events. \label{fig:energy_reconstruction}}
\end{figure}

The reconstruction performance of the primary particle energy for the simulated test set and the generated data sets is shown in \Cref{fig:energy_reconstruction}.
The reconstruction works well over the whole energy, showing that the energy is correctly represented in the generated IACT images.
In direct comparison, the energy of the generated data set shows a slightly better reconstruction performance than the simulated test data, which is mainly visible through the better energy resolution.
Considering  the generator receives feedback from the energy constrainer for the enforced implementation of the energy label, this indicates that the generator is likely underestimating the fluctuations for a given energy, even being able to generate events from the bulk of the distribution successfully.

The few outliers for the energy reconstruction, particularly seen in the simulated data set, are images in which the Cherenkov light of the shower did not trigger the telescopes but single muons, which could not be rejected in the used analysis setup and are represented by a ring-like signal on the camera image~\cite{chalmecalvet2014muon}.
Interestingly, we do not find similar examples in the generated data set.
Due to the small number of muon-like images in the simulated data set --- a handful of muon images of around 10 to 20 out of 500,000 gamma images --- the generator cannot properly learn to generate such images using the given data.
The very small number of outliers (<10) present for the energy reconstruction of the generated data is caused by the generator mapping the input information to a non-realistic camera image that, however, could be rejected using a dedicated cut.

\begin{figure}[t!]
\centering
\includegraphics[width=.99\textwidth]{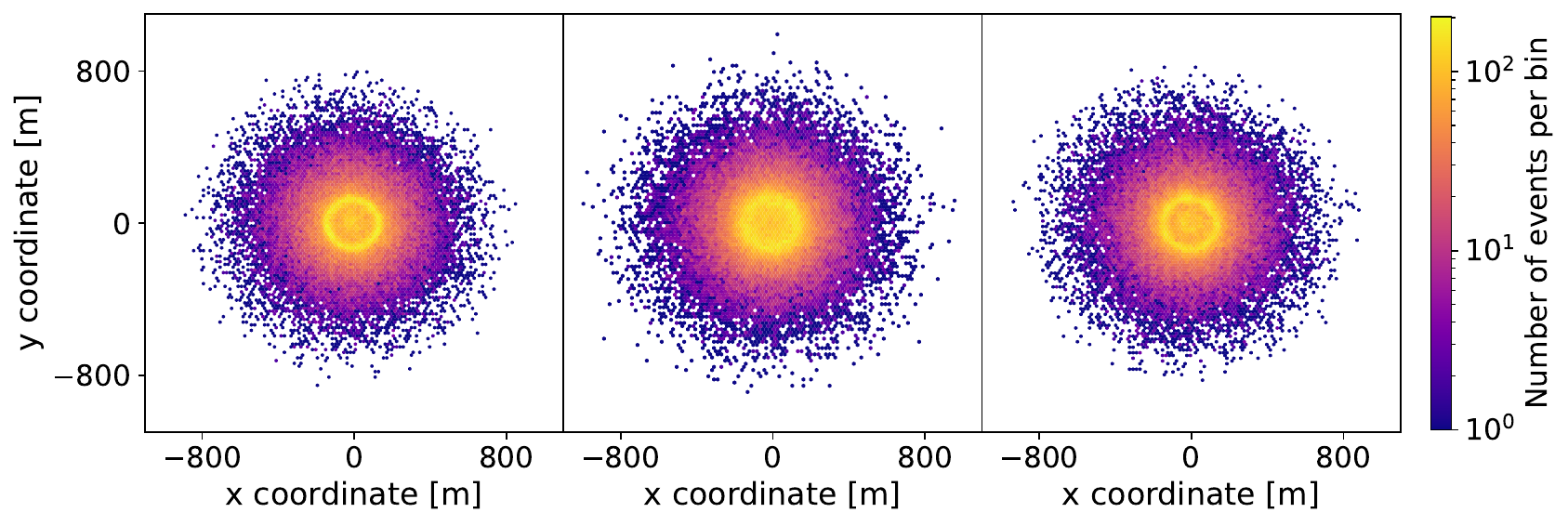}
\caption{Reconstruction of the \(x\) and \(y\) coordinate of the air shower impact point \(I\) using the impact constrainer network for simulated (left) and generated (right) events. In the middle, the true impact points of the simulated events, which are also used for the generation of events, are shown. \label{fig:impact_point_map}}
\end{figure}

In \Cref{fig:impact_point_map}, the reconstruction of the air shower impact point for simulated (left) and generated (right) images are shown using the constrainer network and compared to the MC truth (middle).
For both cases, a larger density of impact points close to the telescope is visible due to the fact that only high-energy showers trigger the telescope from further distances.
The visible ring with a diameter of roughly $150$~m represents the typical radius of a Cherenkov cone for a vertical shower at the H.E.S.S. site.
Further, more events are reconstructed in the array's center for the simulation and the generated images.
This visualizes a slight bias of the reconstruction, causing a more likely estimate in the center of the array if the constrainer network is not able to infer which direction to reconstruct the respective impact.
This effect is commonly referred to as \emph{regression to the mean}.
The finding that this effect is slightly more pronounced for simulated events likely indicates, similar as found for the energy, that the WGAN is not able to precisely model all fluctuations w.r.t. the input point in the image.
Overall, the reconstruction of the primary particle energy \(E\) and the air shower impact point \(I\) works well showing that the respective physical characteristics are represented in the generated IACT images.

\subsection{Hillas parameters}
\label{sec:hillas_parameters}
Even though generating realistic-looking IACT images conditioned on physical properties is a complex task by itself, a quantitative follow-up analysis of the generated images is required.
In the standard analysis of IACT images, the Hillas parameterization --- a high-level parameterization of the image shape --- is used to reconstruct the arrival direction and energy of the showers as well as to gain information on the primary particle type.
Since the parameterization enables a precise event reconstruction and relies on image characteristics, the Hillas parameters can be excellently employed for the purpose of examining the quality of generated images. 

\begin{figure}[t!]
\begin{subfigure}[b]{0.5\textwidth}
\centering
\includegraphics[width=.65\textwidth]{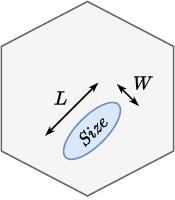}
\label{fig:hillas_lw}
\end{subfigure}
\begin{subfigure}[b]{0.5\textwidth}
\centering
\includegraphics[width=.65\textwidth]{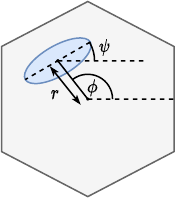}
\label{fig:hillas_angles}
\end{subfigure}
\caption{Illustration of the six investigated Hillas parameters modeling the measured signal as an ellipsis. Left: Hillas size (signals integrated over the full image), length $L$, and width $W$. Right: Radial $r$ and polar $\phi$ coordinate and rotation angle $\psi$. \label{fig:hillas_scheme}}
\end{figure}

To derive the Hillas parameters, as a first step, image cleaning is applied to remove signal pixels containing a large fraction of the night sky background.
Since we used CT5, we are applying the same 9/16 cleaning procedure as performed by the H.E.S.S. collaboration.
We further set a threshold cut (after cleaning) of 250\,p.e. and discard events below the threshold.
For yielding the same number of events for testing after cleaning, we post-generate the required number of events using the WGAN, which amounts to a few percent.
In the second step, the Hillas parameters are derived using ctapipe~\cite{ctapipe-icrc-2021} (v0.19.0 \cite{karl_kosack_2023_7788918}), modeling the Cherenkov light in the camera after cleaning as an ellipsis using the position and values of the pixels containing a signal. In this analysis, we focus on the most important Hillas parameters: intensity, length, width, radial coordinate, polar coordinate, and rotation angle.
An illustration of these six parameters can be found in \Cref{fig:hillas_scheme}.

\begin{figure}[t!]
    \begin{subfigure}[c]{0.45\textwidth}
        \centering
        \includegraphics[width=.99\textwidth]{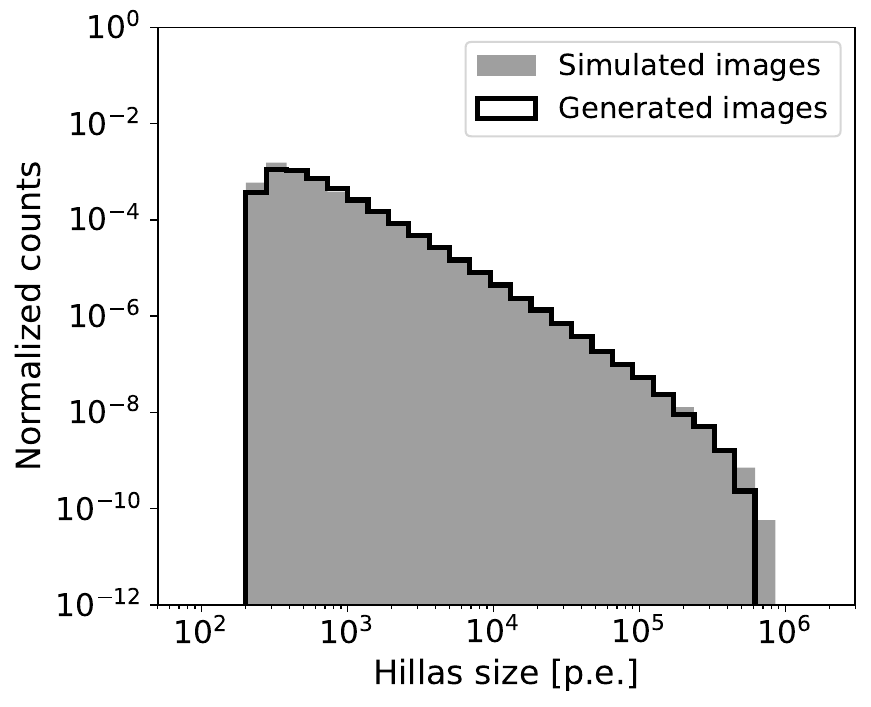}
        \caption{}
        \label{fig:hillas_size}
    \end{subfigure}
    \begin{subfigure}[c]{0.45\textwidth}
        \centering
        \includegraphics[width=.99\textwidth]{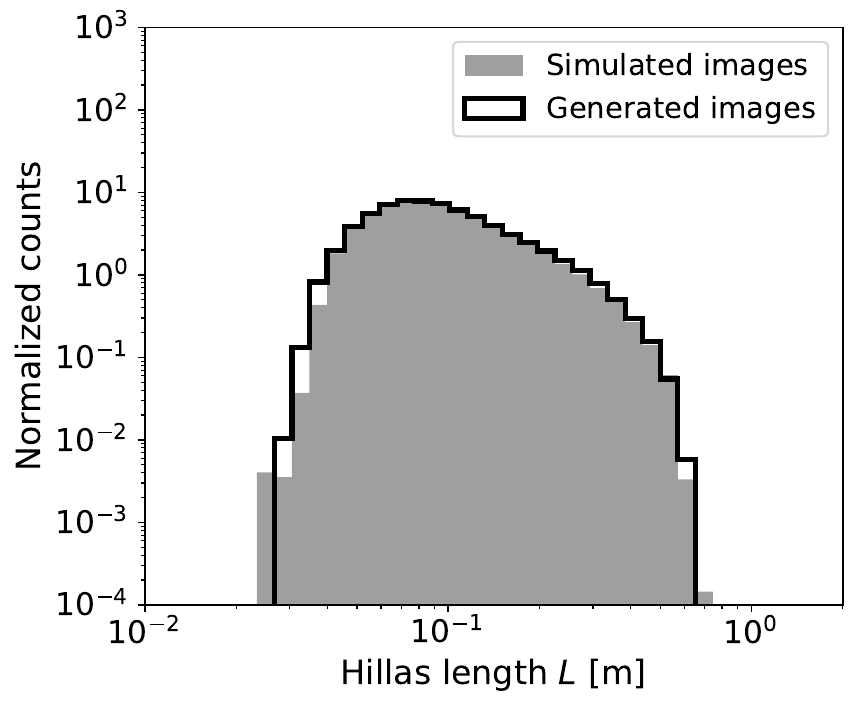} %
        \caption{}
        \label{fig:hillas_length}
    \end{subfigure}
    \begin{subfigure}[c]{1.\textwidth}
        \centering
        \includegraphics[width=.45\textwidth]{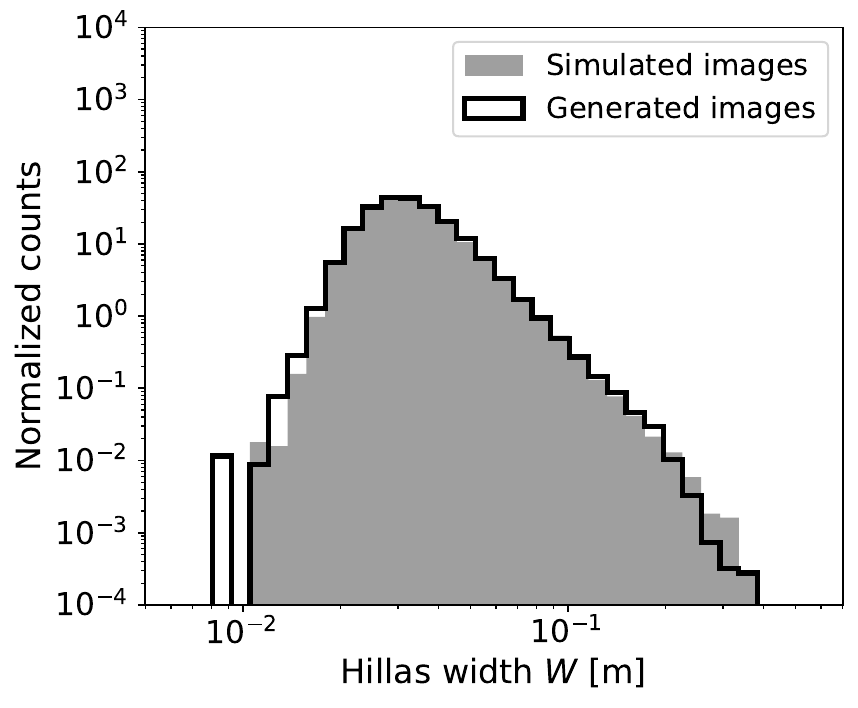}
        \caption{}
        \label{fig:hillas_width}
    \end{subfigure}
\caption{Comparison of the distributions of the Hillas parameters (a) size, (b) length, and (c) width for the generated (black) and simulated (grey) images. The standard 9/16 tail cuts cleaning was applied to both data sets. \label{fig:hillas_params_01}}
\end{figure}

\paragraph{Intensity}
The Hillas intensity (size), shown in \Cref{fig:hillas_size} for both data sets, is defined by integrating the remaining pixel intensities after cleaning and strongly depends on the particle energy and the distance of the telescope to the shower core.
Both low particle energies and shower cores far away from the telescope typically result in small image sizes due to the little Cherenkov light being detected.
On the contrary, large image sizes are obtained when the primary particle has high energy or the shower core is close to the telescope.
The cut-off at low intensities is caused by the size cut described above.
With a decreasing trend to higher values, the sizes reach up to \(10^6\,\text{p.e.}\).
The generated and simulated image distribution shows a very good agreement over the whole range, comprising more than three orders of magnitude.
Only at the lowest and highest intensities, small deviations are visible.
However, at the right tail of the size distribution, the statistics are low.

\begin{figure}[t!]
    \begin{subfigure}[c]{0.45\textwidth}
        \centering
        \includegraphics[width=.99\textwidth]{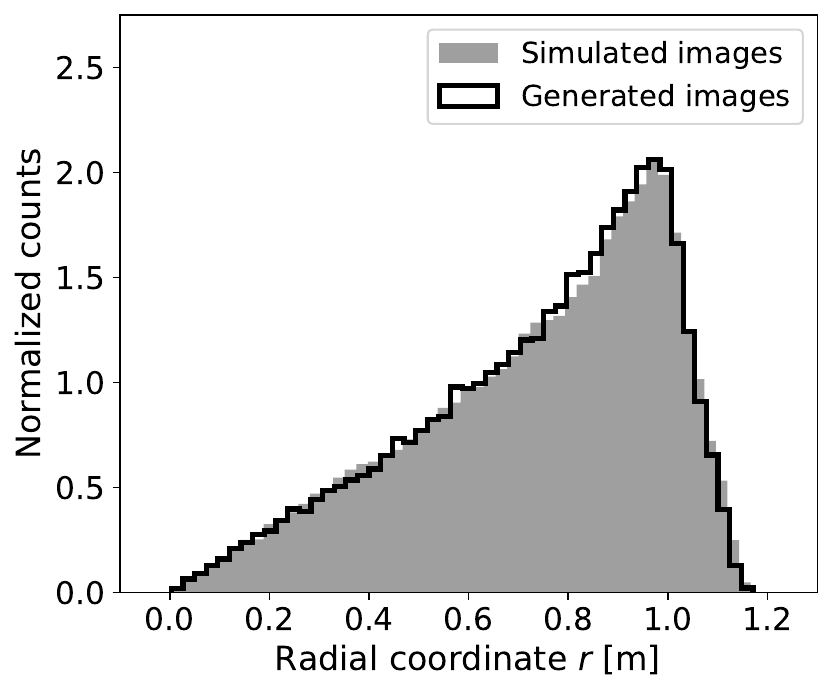}
        \caption{}
        \label{fig:hillas_distance}
    \end{subfigure}
    \begin{subfigure}[c]{0.45\textwidth}
        \centering
        \includegraphics[width=.99\textwidth]{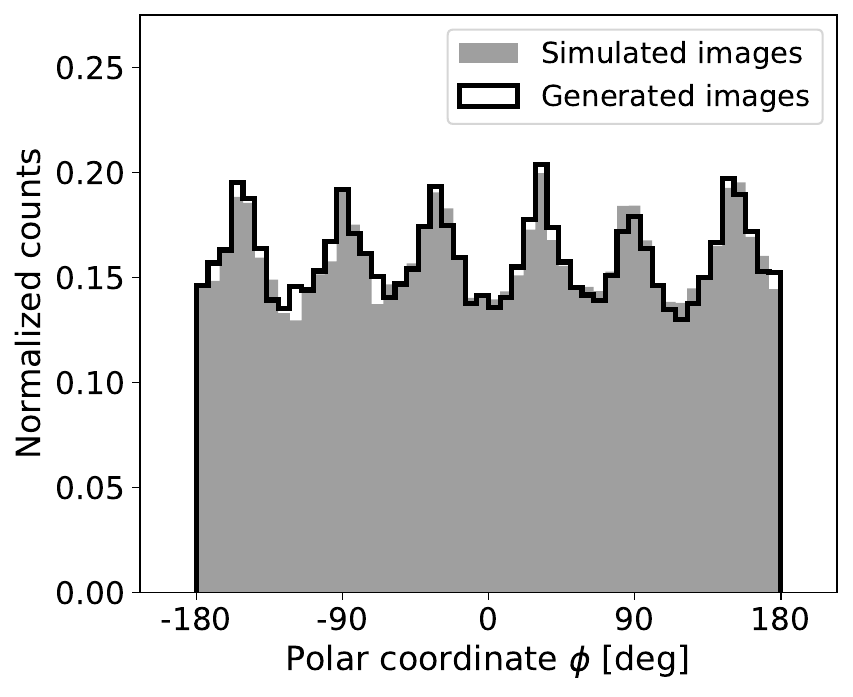}
        \caption{}
        \label{fig:hillas_phi}
    \end{subfigure}
    \begin{subfigure}[c]{1.\textwidth}
        \centering
        \includegraphics[width=.45\textwidth]{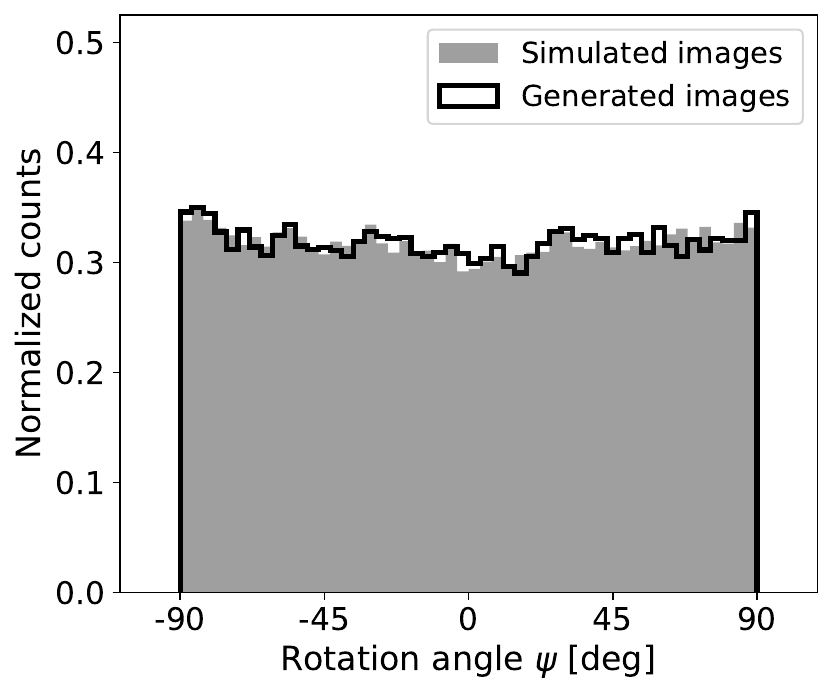}
        \caption{}
        \label{fig:hillas_psi}
    \end{subfigure}
\caption{Comparison of the distributions of the Hillas parameters (a) radial coordinate, (b) polar coordinate, and (c) rotation angle for the generated (black) and simulated (grey) events. The standard 9/16 tail cuts cleaning was applied to both data set.\label{fig:hillas_params}}
\end{figure}

\paragraph{Length and width of the ellipses}
The Hillas length \(L\) and width \(W\) characterize the shape of the signal and are the standard deviations of the measured signal along the major and minor axis of the ellipsis on the camera image, respectively; i.e., they are estimated by taking the square root of the two eigenvalues obtained after estimating the two largest components using a principal component analysis of the image.
The distributions for the Hillas length and width match for the simulated and generated images quite well are shown in \Cref{fig:hillas_length} and \Cref{fig:hillas_width}.
Whereas the bulk of the distribution can be precisely reproduced by the WGAN, at the edges, small differences remain.
These are, however, statistically not very significant, indicating that the WGAN is able to reproduce the shapes of gamma-ray images correctly.

\begin{figure}[t!]
\centering
    \begin{subfigure}[c]{0.99\textwidth}
        \includegraphics[width=.99\textwidth]{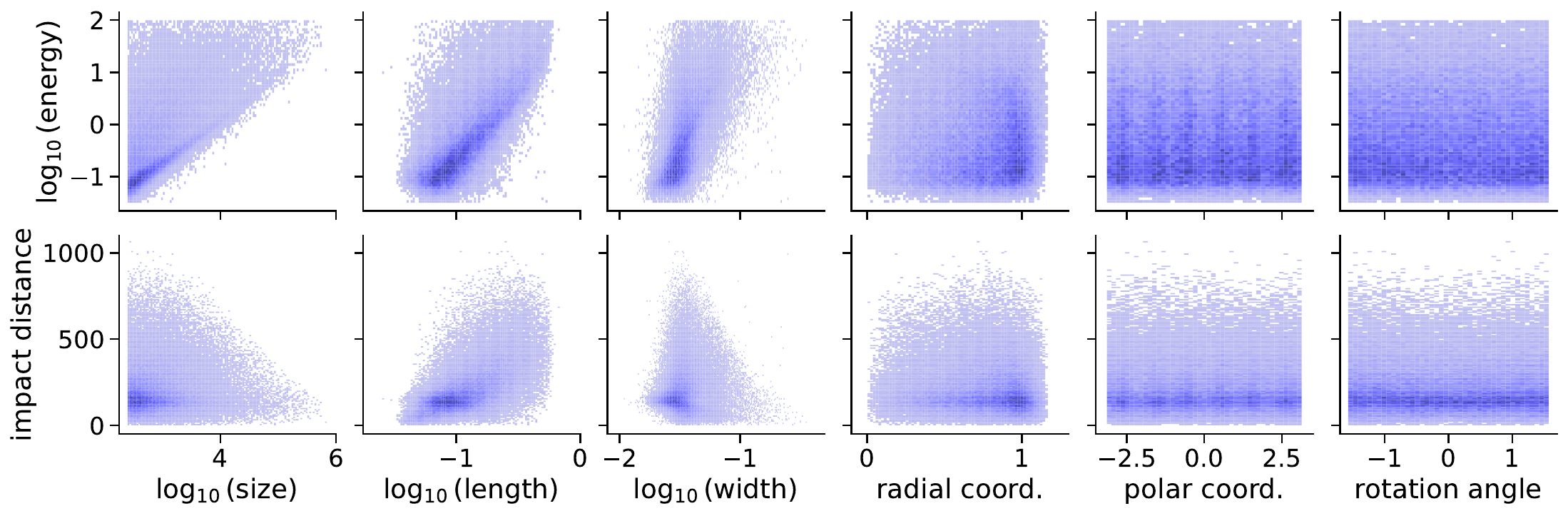}
        \caption{}
        \label{fig:correlations_energy_impact_sim}
    \end{subfigure}
    \begin{subfigure}[c]{0.99\textwidth}
     \includegraphics[width=.99\textwidth]{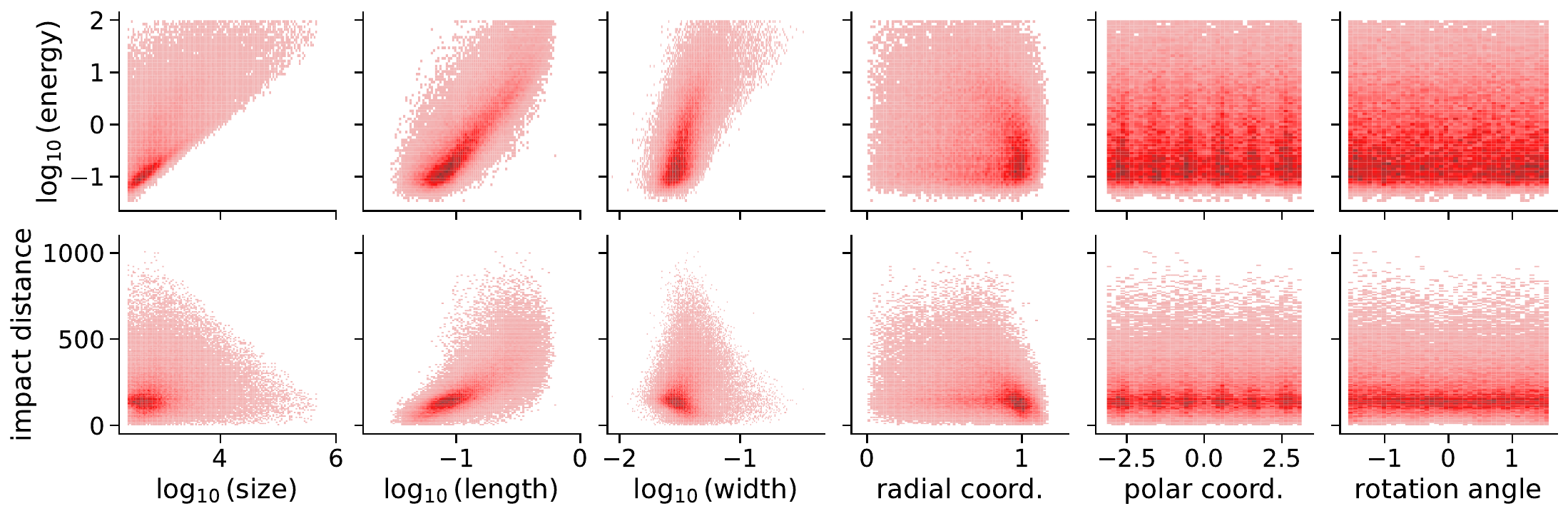}
        \caption{}
        \label{fig:correlations_energy_impact_gen}
    \end{subfigure}
\caption{Correlations between the corresponding physical labels ($E$, and impact distance) and the Hillas parameters for (top) the simulated images shown in blue and (bottom) the image generated using the WGAN shown in red.}
\end{figure}

\paragraph{Rotation and position of the ellipses}
The radial coordinate \(r\) and polar coordinate \(\phi\) describe the location of the center of the ellipsis.
While the radial coordinate, often referred to as local distance, is the distance from the camera center to the ellipsis center, the radial coordinate defines the angle between the \(x-\text{axis}\) of the camera and the ellipsis center.
The parameter distributions for both data sets are shown in \Cref{fig:hillas_distance} and \Cref{fig:hillas_phi}, reaching from 0~m, ellipsis center at the camera center to roughly 1.2~m the edge of the telescope.
Both distributions show a very good agreement, denoting that the whole camera frame is adequately covered by the generated showers of the WGAN, well in line with the finding of a realistic occupancy (cf. \Cref{fig:occupancy_image}).

Additionally, the distributions of the rotation angle $\psi$ of the ellipsis match well, as shown in \Cref{fig:hillas_psi}.
The coordinate is defined as the angle between the \(x-\text{axis}\) on the camera and the major axis of the ellipsis and, therefore, features six maxima due to the hexagonal camera design of FlashCam.
Combined with the previous observations, this finding indicates that many different shower geometries are generated by the algorithm, excluding mode collapsing present in previous approaches~\cite{taiga_2}.

\begin{figure}[t]
\centering
\includegraphics[width=.99\textwidth]{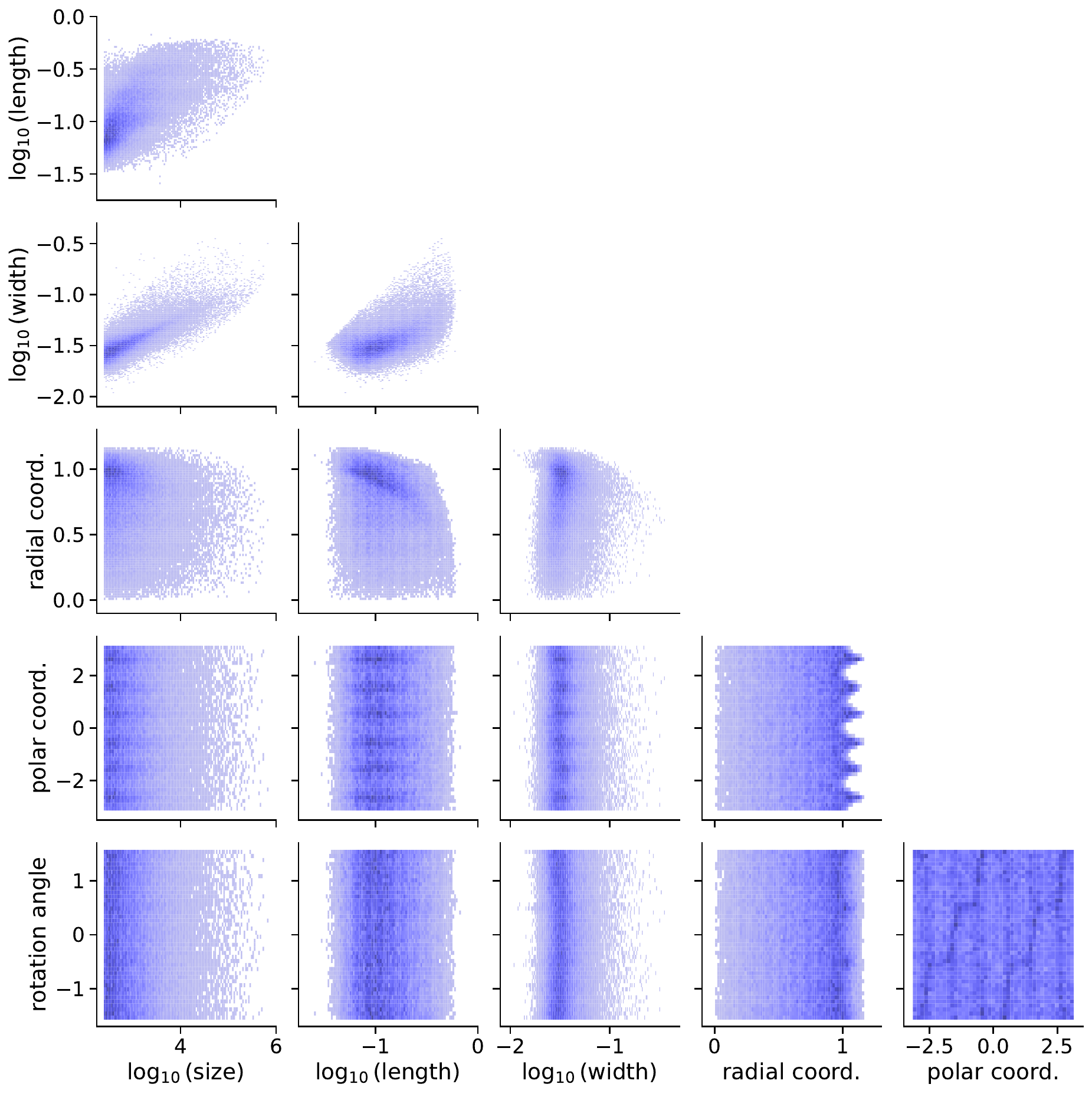}
\caption{Cross-correlation in-between the six investigated Hillas parameters for the simulated images. \label{fig:correlations_hillas_sim}}
\end{figure}

\begin{figure}[th]
\centering
 \includegraphics[width=.99\textwidth]{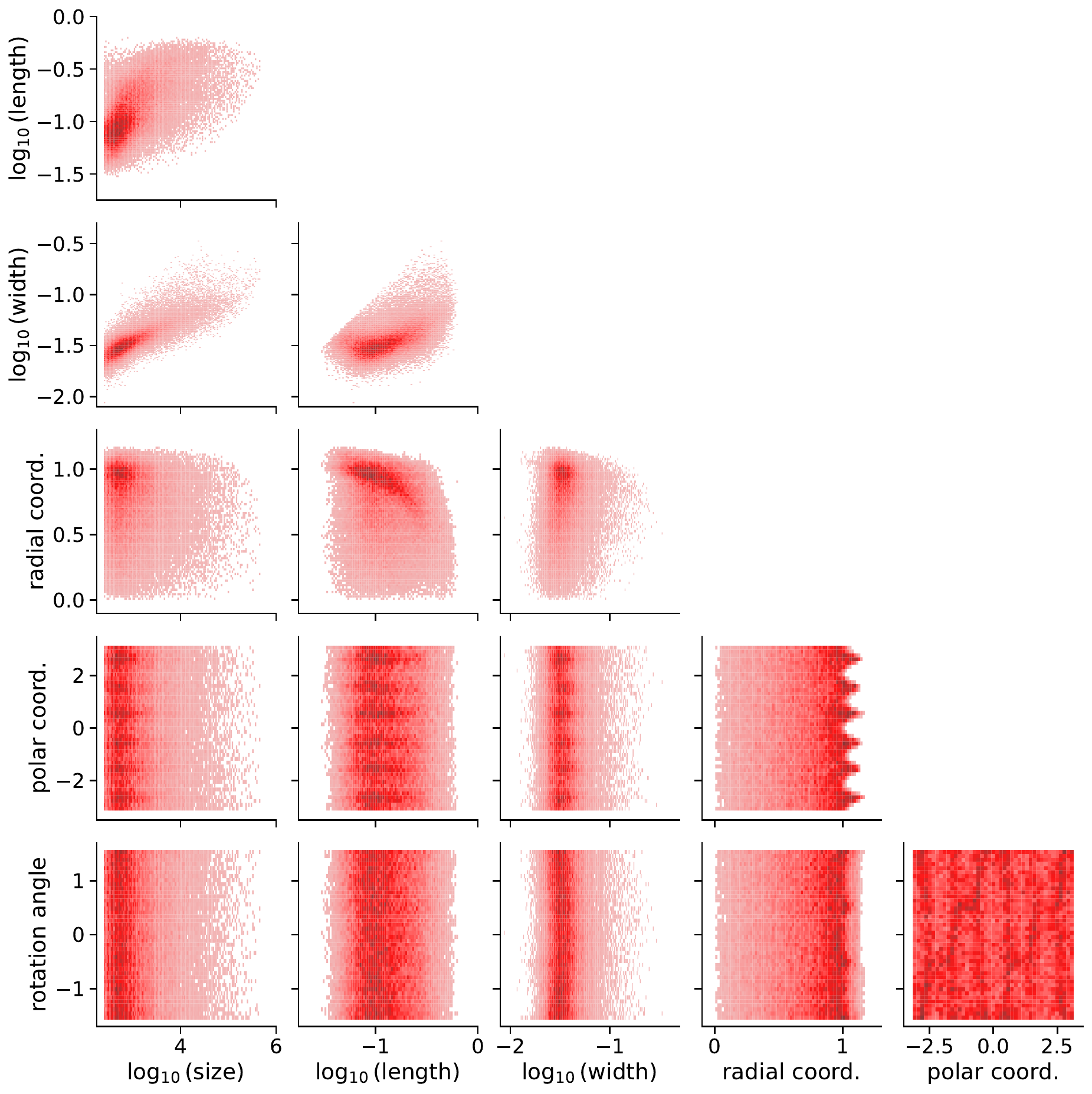}
\caption{Cross-correlation in-between the six investigated Hillas parameters as found in generated images of the WGAN. \label{fig:correlations_hillas_gen}}
\end{figure}

\subsection{Correlation of high-level parameters}
\label{sec:parameter_correlations}

Lastly, the correlation of the high-level parameters, so the energy, the impact point, and the Hillas parameters, is investigated to verify that the generator is able to reproduce the most complex relations regarding IACT images.
Since the images and their characteristics mostly depend on the distance of the telescope to the impact point, we investigate $I = \sqrt{I_x^2 + I_y^2}$ hereafter.

After studying the implementation of the physics conditions during the image generation using the two constrainer networks as discussed in \Cref{sec:energy_and_impact} by reconstructing the labels, we now investigate their correlation with high-level parameters.
The corresponding correlations for simulated and generated images are shown in \Cref{fig:correlations_energy_impact_sim} and \Cref{fig:correlations_energy_impact_gen}, in which simulated events are shown in blue and events generated using the WGAN in red.
When comparing both Figures, the inter-dependencies of each of the various Hillas parameters are very diverse.
However, these very different correlations are qualitatively similar in both the simulated and generated data sets.
In particular, the bulk of the distribution is well reproduced for all cases.
Differences remain, especially for the length and the local distance.
These identified differences might be related to the small but still noteworthy performance difference discussed in Sec.~\ref{sec:energy_and_impact}.
Further changes in the training strategy, e.g., by enhancing the quality of the feedback of the constrainer networks using spectral normalization~\cite{miyato2018spectral}, could possibly improve the performance.
Nonetheless, the overall correlations are reproduced in much detail.

Next, the correlations between the Hillas parameters themselves are examined.
The parameter correlations of the simulated images are shown in \Cref{fig:correlations_hillas_sim} in blue, while the correlations using the images generated with the WGAN are shown in red in \Cref{fig:correlations_hillas_gen}.
The overall correlations seem to be very well-produced using the WGAN, and the covered phase space matches, i.e., no outliers are visible in the generated events.
Also, in the case where no strong correlations are expected, e.g., for the radial coordinates and the rotation angle, the WGAN does not induce artificial correlations not present in the simulated test data set.
Only very minor differences remain, e.g., in the correlation between length and size.

Conclusively, we have demonstrated that the WGAN is not only able to reproduce the 1-dimensional distributions of low- and high-level parameters with good quality but also shows high fidelity with respect to their correlations.
This is a significant improvement in comparison to previous work since not only IACT images with a high resolution of more than 1,500 pixels without mode collapsing were generated, but also the generated images were comprehensively analyzed using low-level and image parameters and their correlation.
The finding that IACT images can be generated with high fidelity using deep generative models opens up promising opportunities for fast and efficient simulations in the era of CTA.

\section{Towards deep-learning-based image generation for IACT arrays}
In our work, we demonstrate the first successful generation of realistic-looking single telescope camera images with high fidelity using generative models.
Modern observatories, however, feature arrays of IACTs, requiring adaptations of the algorithm that need to be investigated further.
Building upon the current progress, several key enhancements --- stereoscopic image generation, the inclusion of NSB, and the conditioning of a wider range of physical and configuration parameters --- are essential to provide comprehensive and reliable simulations.

\paragraph{Stereoscopic aspect of air shower image generation}

The key design of modern observatories like H.E.S.S., VERITAS, or CTA is the possibility of obtaining a stereoscopic view of the detected air showers by performing observations using arrays of telescopes.
The additional information gained through stereoscopic observations results in a significant improvement in the reconstruction of air showers and, thus, more accurate information about the primary particle and its origin.
Consequently, air shower simulations of any kind need to be able to produce camera images with an accurate representation of stereoscopy.
The successful implementation of stereoscopy into deep-learning-based algorithms is an ongoing challenge in the community, mostly manifesting itself in shortcomings of the energy and angular reconstruction~\cite{Shilon_2019, Jacquemont_2021, Brill_2019, nieto_hex, ct_learn, Spencer_2021}.
Although the presented WGAN approach offers the possibility of generating camera images for several telescopes simultaneously by simply changing the impact parameter for another telescope, stereoscopy is not explicitly enforced.
The concept of attention~\cite{vaswani2023attention}, which showed great progress in extracting global correlations, could offer new possibilities to enforce stereoscopy, e.g., using relation reasoning modules~\cite{hashemi2023ultrahighresolution}, in image generations as well as stereoscopic reconstruction using neural networks.

\paragraph{Inclusion of the generation of NSB light}

Besides transitioning from single telescope images to stereoscopic events, another challenge in deep-learning-based image generation is the accurate description of the NSB.
To date, the generation of NSB has not been investigated using generative models.
A common method of simulating NSB light is the usage of the \(nsb\) python package \cite{nsb_package} or an extended version of the package described in Ref.~\cite{spencer2023advanced}.
To establish an entire fast simulation chain for IACTs, we intend to add NSB using a second generative model trained separately using measurements of the NSB.
The final image is obtained by combining the air shower image and the NSB image, ensuring the ability to unambiguously discriminate, even after the generation, the respective contribution of NSB and the air shower at the pixel level

\paragraph{Conditioning on additional physical parameters and observation conditions}

While the image generation in this work is already conditioned on two very important physical parameters related to IACT events, there are limitations to the variation of different showers that can be generated.
Besides conditioning the generation process on shower properties like $X_\mathrm{max}$ that are not directly accessible with MC codes, an example of another parameter that particularly impacts the detected distribution of Cherenkov light is the zenith angle, which is interconnected to the NSB rate.
To enable the generation of air shower images under different observation conditions, e.g., including zenith angle and atmospheric conditions, updates of the constrainer networks are needed for both the framework for generating the air showers as well as a framework for generating the NSB.

\section{Conclusion}
In this work, we used deep generative models to generate gamma-ray air-shower images measured by Imaging Air Cherenkov Telescopes (IACTs).
The CT5 telescope at H.E.S.S., equipped with a state-of-the-art camera --- holding more than 1,500 pixels --- foreseen for the upcoming CTA flagship project was used in this study.
Our approach utilizes Wasserstein Generative Adversarial Networks (WGAN) and Convolutional Neural Networks to accelerate the computing-intensive simulations of IACT arrays significantly.
Whereas the whole simulation chain consists of the simulation of the air shower, followed by simulating the instrument response, we followed an end-to-end approach, enabling us to generate new gamma-ray images given an air shower energy and the shower impact position on the ground.

For the first time, we have demonstrated the fast generation of high-quality gamma-ray images that feature the conditioned physical properties, whereas the generated events do not only show high fidelity with respect to low and high-level image parameters but also high variety.
We have comprehensively examined the quality of the generated images and found an excellent agreement with simulated images when analyzing low-level parameters, such as the distribution of pixel intensities, signal intensities, and camera occupancy.
We further studied the shapes of the signal patterns in the generated images using the Hillas parameters and their correlation and found a good agreement with simulated images.
Only minor differences limited to phase space regions with low density were found, causing slight under-fluctuations in the generated events, which leave room for improvements.

Future work will focus on the generation beyond single telescope data to facilitate the acceleration of simulations for IACT arrays and the generation of showers induced by protons and other hadrons.
Also, we plan to separate the generation process into the generation of the air shower and the instrument response to make the algorithm versatile and valuable outside the IACT community.
Furthermore, this enables adding different levels of NSB in the event generation process, e.g., by adding a dedicated NSB generative model trained using simulations or measurement data.

The reached simulation speed-up of $10^3$ ($10^5$) on the CPU (GPU), combined with the convincing image quality comprehensively studied within this work, leaves promising prospects for fast, computationally efficient, and sustainable simulations in gamma-ray astronomy.

\clearpage
\newpage

%\appendix
%\section{Some title}
%Please always give a title also for appendices.

\acknowledgments
We thank the H.E.S.S. Collaboration for allowing us to use H.E.S.S. simulations for this publication. We further thank Johannes Schaefer, Simon Steinmassl, and the H.E.S.S. simulation team for running and producing the simulations.
Additionally, we thank Benedetta Bruno for preparing the simulations and valuable discussions, and Samuel T. Spencer for helpful comments on the manuscript.
The authors gratefully acknowledge the scientific support and HPC resources provided by the Erlangen National High Performance Computing Center (NHR@FAU) of the Friedrich-Alexander-Universität Erlangen-Nürnberg (FAU) under the NHR project b129dc.
NHR funding is provided by federal and Bavarian state authorities. NHR@FAU hardware is partially funded by the German Research Foundation (DFG) – 440719683.

% Bibliography

%% [A] Recommended: using JHEP.bst file
\bibliographystyle{JHEP}
\bibliography{biblio.bib}

% \clearpage
% \newpage

\appendix 

\newpage
\section{Appendix}
\begin{figure}[h]
\centering
\includegraphics[width=1.\textwidth]{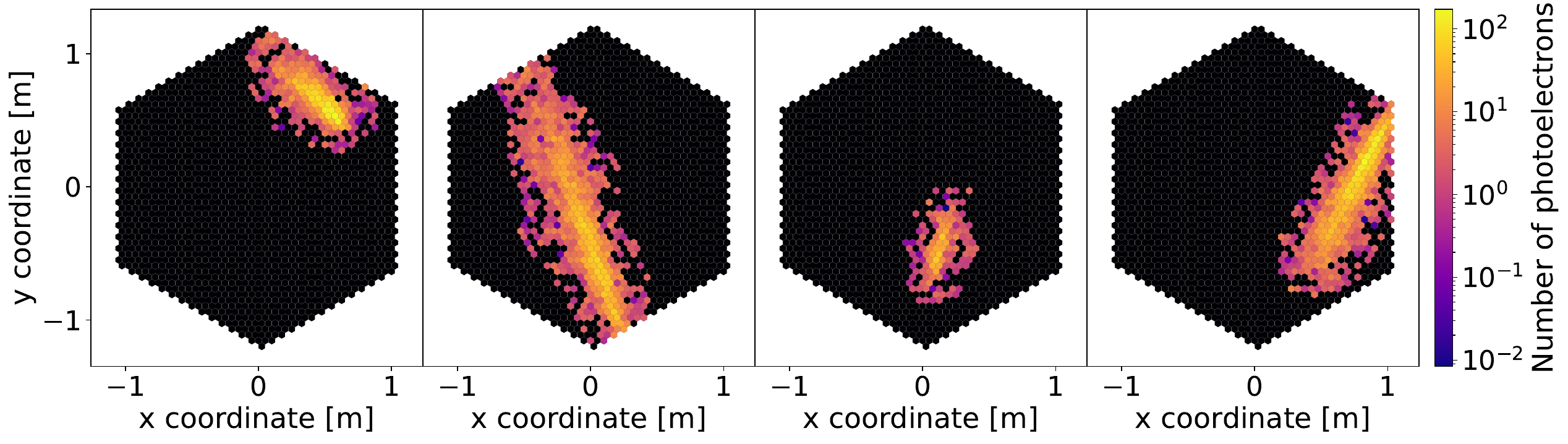}
\qquad
\includegraphics[width=1\textwidth]{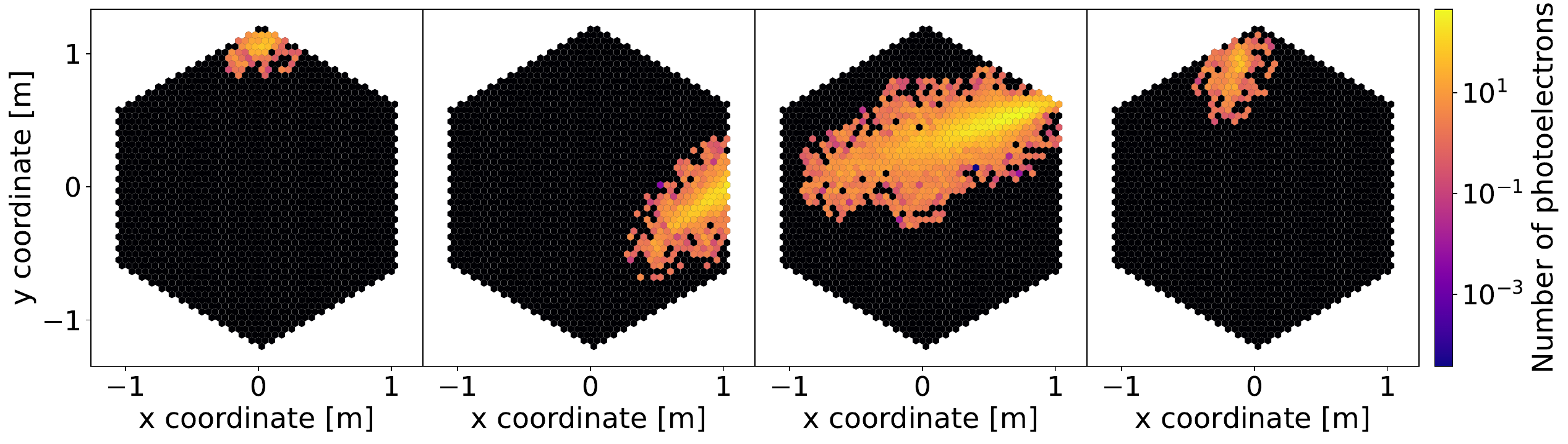}
\qquad
\includegraphics[width=1\textwidth]{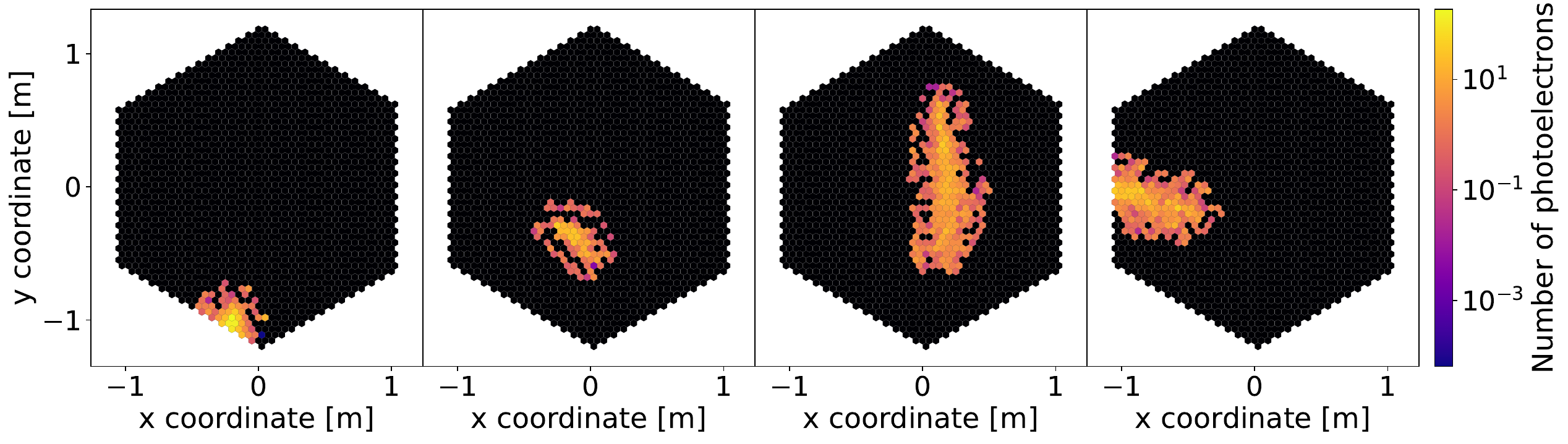}
\qquad
\includegraphics[width=1\textwidth]{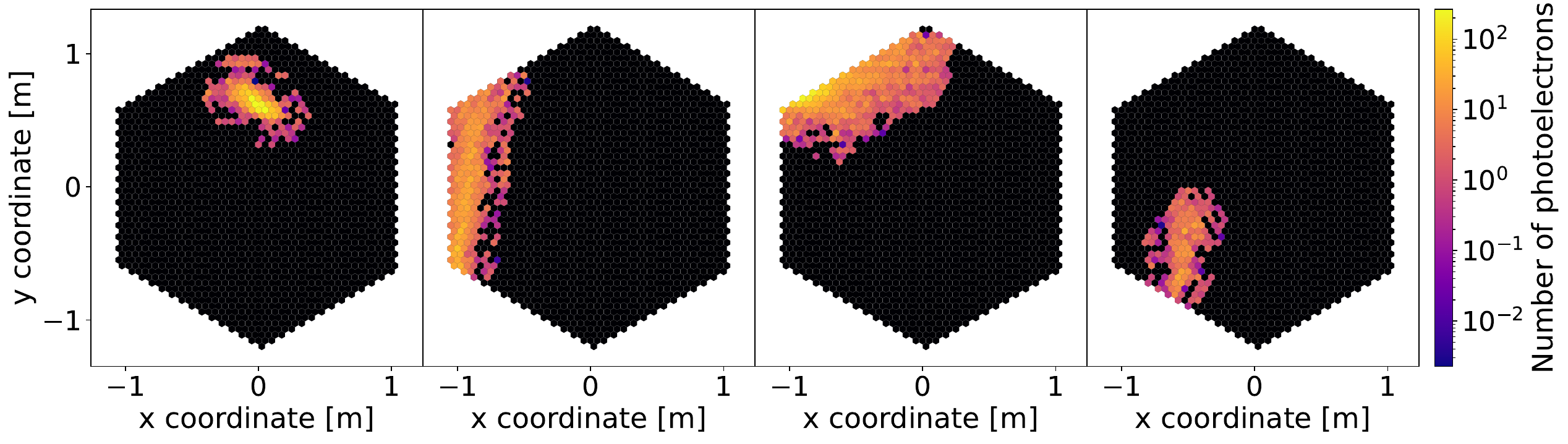}
\caption{In the first two rows, eight random IACT images from the simulated test data set are shown. The IACT images in the last two rows are randomly chosen from the generated data set.
\label{fig:add_images}}
\end{figure}

\paragraph{Generator}
The generator architecture is shown in Table~\ref{tab:gen_arch}. 
To enforce the saturation present in the simulations at about 4176 p.e., we clip the final output of the generator network using a ReLU in the final layer. We further mask all pixels in the last layer to ensure that the generator only generates signals in the camera and not the padding pixels, which are needed to represent the camera and its hexagonal design in a 2D cartesian image.
As a post-processing function, we are using extended 4/7 cleaning.
%and a cutoff at 250 p.e.

\begin{table}[h!]
\centering
\begin{centering}
\begin{tabular}{ c c c c c }
%\multicolumn{5}{l}{Generator architecture} \\
\hline\hline
No & Input shape & Layer & Output shape & Features \\ 
\hline
1 & 100 + 1 + 2 & Dense + ReLU & 12544 & -- \\
2 & 12544 & Reshape & $7 \times 7$ & 256 \\
3 & $7 \times 7$ & Conv2D + ReLU & $7 \times 7 $ & 256 \\
4 & $7 \times 7$ & Conv2D + ReLU & $7 \times 7 $ & 256 \\
5 & $7 \times 7$  & Upsampling & $14 \times 14 $ & 256 \\
6 & $14 \times 14 $ & Conv2D + ReLU & $14 \times 14 $ & 128 \\
7 & $14 \times 14 $ & Conv2D + ReLU & $14 \times 14 $ & 128 \\
8 & $14 \times 14 $   & Upsampling & $28 \times 28$ & 128 \\
9 & $28 \times 28$ & Conv2D + ReLU & $28 \times 28$ & 64 \\
10 & $28 \times 28$ & Conv2D + ReLU & $28 \times 28$ & 64 \\
11 & $28 \times 28 $   & Upsampling & $56 \times 56$ & 64 \\
12 & $56 \times 56$ & Conv2D + ReLU & $56 \times 56$ & 32 \\
13 & $56 \times 56$ & Conv2D + ReLU & $56 \times 56$ & 32 \\
14 & $56 \times 56$ & Conv2D + ReLU & $56 \times 56$ & 1 \\
15 & $56 \times 56$ & Multiply & $56 \times 56$ & -- \\
\hline
\end{tabular}
\caption{Architecture of the generator network.}
\label{tab:gen_arch}
\end{centering}
\end{table}

\begin{table}[h!]
\centering
\begin{centering}
\begin{tabular}{ c c c c c c }
%\multicolumn{5}{l}{Critic architecture} \\
\hline\hline
No & Input shape & Layer & Output shape & Features \\ 
\hline
1 & $56 \times 56$ & Conv2D + LeakyReLU & $56 \times 56$ & 32 \\
\hline
2 & $56 \times 56$ & Conv2D + LeakyReLU & $56 \times 56$ & 32 & \multirow{4}{*}{$ 2\,\times $}\\
3 & $56 \times 56$ & Conv2D & $56 \times 56$ & 32 & \\
4 & $56 \times 56$ & Add & $56 \times 56$ & 32 & \\
5 & $56 \times 56$ & LeakyReLU & $56 \times 56$ & 32 & \\
\hline
6 & $56 \times 56$ & AvgPooling & $28 \times 28$ & -- \\
7 & $28 \times 28$ & Conv2D + LeakyReLU & $28 \times 28$ & 64 \\
\hline
8 & $28 \times 28$ & Conv2D + LeakyReLU & $28 \times 28$ & 64 & \multirow{4}{*}{$ 2\,\times $}\\
9 & $28 \times 28$ & Conv2D & $28 \times 28$ & 64 & \\
10 & $28 \times 28$ & Add & $28 \times 28$ & 64 & \\
11 & $28 \times 28$ & LeakyReLU & $28 \times 28$ & 64 & \\
\hline
12 & $28 \times 28$ & AvgPooling & $14 \times 14$ & -- \\
13 & $14 \times 14$ & Conv2D + LeakyReLU & $14 \times 14$ & 128 \\
\hline
14 & $14 \times 14$ & Conv2D + LeakyReLU & $14 \times 14$ & 128 & \multirow{4}{*}{$ 2\,\times $}\\
15 & $14 \times 14$ & Conv2D & $14 \times 14$ & 128 & \\
16 & $14 \times 14$ & Add & $14 \times 14$ & 128 & \\
17 & $14 \times 14$ & LeakyReLU & $14 \times 14$ & 128 & \\
\hline
18 & $14 \times 14$ & AvgPooling & $7 \times 7$ & -- \\
19 & $7 \times 7$ & Conv2D + LeakyReLU & $7 \times 7$ & 256 \\
\hline
20 & $7 \times 7$ & Conv2D + LeakyReLU & $7 \times 7$ & 256 & \multirow{4}{*}{$ 2\,\times $}\\
21 & $7 \times 7$ & Conv2D & $7 \times 7$ & 256 & \\
22 & $7 \times 7$ & Add & $7 \times 7$ & 256 & \\
23 & $7 \times 7$ & LeakyReLU & $7 \times 7$ & 256 & \\
\hline
24 & $7 \times 7$ & Flatten & 12544 & -- \\
25 & 12544 & Dense & 100 & -- \\
26 & 1 & Dense & 50 & -- \\
27 & 2 & Dense & 50 & -- \\
28 & 100+50+50 & Concatenate & 200 & -- \\
29 & 200 & Dense + LeakyReLU & 100 & -- \\
30 & 100 & Dense + LeakyReLU & 100 & -- & \\
31 & 100 & Dense & 100 & -- & \\
32 & 100 & Add & 100 & -- & \\
33 & 50 & Dense & 1 & -- \\
\hline
\end{tabular}
\caption{Architecture of the critic network.}
\label{tab:critic_arch}
\end{centering}
\end{table}

\begin{table}[h!]
\centering
\begin{centering}
\begin{tabular}{ c c c c c }
%\multicolumn{5}{l}{Energy constrainer architecture} \\
\hline\hline
No & Input shape & Layer & Output shape & Features \\ 
\hline
1 & $56 \times 56$ & Conv2D + LeakyReLU & $56 \times 56$ & 16 \\
2 & $56 \times 56$ & Conv2D + LeakyReLU & $56 \times 56$ & 16 \\
3 & $56 \times 56$ & Conv2D & $56 \times 56$ & 16 \\
4 & $56 \times 56$ & Add (1+3) & $56 \times 56$ & 16 \\
5 & $56 \times 56$ & LeakyReLU & $56 \times 56$ & 16 \\
5 & $56 \times 56$ & AvgPooling & $28 \times 28$ & -- \\
6 & $28 \times 28$ & Conv2D + LeakyReLU & $28 \times 28$ & 16 \\
7 & $28 \times 28$ & Conv2D + LeakyReLU & $28 \times 28$ & 16 \\
8 & $28 \times 28$ & Conv2D & $28 \times 28$ & 16 \\
9 & $28 \times 28$ & Add (6+8) & $28 \times 28$ & 16 \\
10 & $28 \times 28$ & LeakyReLU & $28 \times 28$ & 16 \\
11 & $28 \times 28$ & AvgPooling & $14 \times 14$ & -- \\
12 & $14 \times 14$ & Conv2D + LeakyReLU & $14 \times 14$ & 32 \\
13 & $14 \times 14$ & Conv2D + LeakyReLU & $14 \times 14$ & 32 \\
14 & $14 \times 14$ & Conv2D & $14 \times 14$ & 32 \\
15 & $14 \times 14$ & Add (12+14) & $14 \times 14$ & 32 \\
16 & $14 \times 14$ & LeakyReLU & $14 \times 14$ & 32 \\
17 & $14 \times 14$ & AvgPooling & $7 \times 7$ & -- \\
18 & $7 \times 7$ & Conv2D + LeakyReLU & $7 \times 7$ & 32 \\
19 & $7 \times 7$ & Conv2D + LeakyReLU & $7 \times 7$ & 32 \\
20 & $7 \times 7$ & Conv2D & $7 \times 7$ & 32 \\
21 & $7 \times 7$ & Add (18+20) & $7 \times 7$ & 32 \\
22 & $7 \times 7$ & LeakyReLU & $7 \times 7$ & 32 \\
23 & $7 \times 7$ & AvgPooling & $3 \times 3$ & -- \\
24 & $3 \times 3$ & Conv2D + LeakyReLU & $3 \times 3$ & 64 \\
25 & $3 \times 3$ & Conv2D + LeakyReLU & $3 \times 3$ & 64 \\
26 & $3 \times 3$ & Conv2D & $3 \times 3$ & 64 \\
27 & $3 \times 3$ & Add (18+20) & $3 \times 3$ & 64 \\
28 & $3 \times 3$ & LeakyReLU & $3 \times 3$ & 64 \\
29 & $3 \times 3$ & Flatten & 576 & 64 \\
30 & 576 & Dense + LeakyReLU & 50 & 64 \\
31 & 2 & Dense + LeakyReLU & 50 & -- \\
32 & 50+50 & Concatenate (30+31) & 100 & -- \\
33 & 100 & Dense + LeakyReLU & 50 & -- \\
34 & 50 & Dropout & 50 & -- \\
35 & 50 & Dense & 1 & -- \\
\hline
\end{tabular}
\caption{Architecture of the energy constrainer network.}
\label{tab:energy_arch}
\end{centering}
\end{table}

\begin{table}[h!]
\centering
\begin{centering}
\begin{tabular}{ c c c c c }
%\multicolumn{5}{l}{Impact constrainer architecture} \\
\hline\hline
No & Input shape & Layer & Output shape & Features \\ 
\hline
1 & $56 \times 56$ & Conv2D + LeakyReLU & $56 \times 56$ & 16 \\
2 & $56 \times 56$ & Conv2D + LeakyReLU & $56 \times 56$ & 16 \\
3 & $56 \times 56$ & Conv2D & $56 \times 56$ & 16 \\
4 & $56 \times 56$ & Add (1+3) & $56 \times 56$ & 16 \\
5 & $56 \times 56$ & LeakyReLU & $56 \times 56$ & 16 \\
5 & $56 \times 56$ & AvgPooling & $28 \times 28$ & -- \\
6 & $28 \times 28$ & Conv2D + LeakyReLU & $28 \times 28$ & 16 \\
7 & $28 \times 28$ & Conv2D + LeakyReLU & $28 \times 28$ & 16 \\
8 & $28 \times 28$ & Conv2D & $28 \times 28$ & 16 \\
9 & $28 \times 28$ & Add (6+8) & $28 \times 28$ & 16 \\
10 & $28 \times 28$ & LeakyReLU & $28 \times 28$ & 16 \\
11 & $28 \times 28$ & AvgPooling & $14 \times 14$ & -- \\
12 & $14 \times 14$ & Conv2D + LeakyReLU & $14 \times 14$ & 32 \\
13 & $14 \times 14$ & Conv2D + LeakyReLU & $14 \times 14$ & 32 \\
14 & $14 \times 14$ & Conv2D & $14 \times 14$ & 32 \\
15 & $14 \times 14$ & Add (12+14) & $14 \times 14$ & 32 \\
16 & $14 \times 14$ & LeakyReLU & $14 \times 14$ & 32 \\
17 & $14 \times 14$ & AvgPooling & $7 \times 7$ & -- \\
18 & $7 \times 7$ & Conv2D + LeakyReLU & $7 \times 7$ & 32 \\
19 & $7 \times 7$ & Conv2D + LeakyReLU & $7 \times 7$ & 32 \\
20 & $7 \times 7$ & Conv2D & $7 \times 7$ & 32 \\
21 & $7 \times 7$ & Add (18+20) & $7 \times 7$ & 32 \\
22 & $7 \times 7$ & LeakyReLU & $7 \times 7$ & 32 \\
23 & $7 \times 7$ & AvgPooling & $3 \times 3$ & -- \\
24 & $3 \times 3$ & Conv2D + LeakyReLU & $3 \times 3$ & 64 \\
25 & $3 \times 3$ & Conv2D + LeakyReLU & $3 \times 3$ & 64 \\
26 & $3 \times 3$ & Conv2D & $3 \times 3$ & 64 \\
27 & $3 \times 3$ & Add (18+20) & $3 \times 3$ & 64 \\
28 & $3 \times 3$ & LeakyReLU & $3 \times 3$ & 64 \\
29 & $3 \times 3$ & Flatten & 576 & 64 \\
30 & 576 & Dense + LeakyReLU & 50 & 64 \\
31 & 1 & Dense + LeakyReLU & 50 & -- \\
32 & 50+50 & Concatenate (30+31) & 100 & -- \\
33 & 100 & Dense + LeakyReLU & 50 & -- \\
34 & 50 & Dropout & 50 & -- \\
35 & 50 & Dense & 2 & -- \\
\hline
\end{tabular}
\caption{Architecture of the impact constrainer network.}
\label{tab:impact_arch}
\end{centering}
\end{table}

\clearpage

\end{document}